\documentclass{aa}  

\usepackage{graphicx}
\usepackage{txfonts}
\usepackage{amsmath}
\usepackage{amssymb}
\usepackage{natbib}

\graphicspath{{./figures/}}
\bibpunct{(}{)}{;}{a}{}{,}
\newlength{\minuslength}
\begin{document}

   \title{Modeling chemistry during star formation: Water deuteration in dynamic star-forming regions}


   \author{S. S. Jensen\inst{1}\thanks{\email{sigurd.jensen@nbi.ku.dk}} 
   \and J. K. J{\o}rgensen\inst{1}
   \and K. Furuya\inst{2}
   \and T. Haugb{\o}lle\inst{1}
   \and Y. Aikawa\inst{3} }

   \institute{Niels Bohr Institute \& Centre for Star and Planet Formation, University of Copenhagen, {\O}ster Voldgade 5-7, DK-1350 Copenhagen K, Denmark
   \and National Astronomical Observatory of Japan, Osawa 2--21--1, Mitaka, Tokyo 181--8588, Japan
   \and Department of Astronomy, The University of Tokyo, Tokyo, 113-0033, Japan}

   \date{Draft date: \today}

 
  \abstract
   {Recent observations of the HDO/H$_2$O ratio toward protostars in isolated and clustered environments show an apparent dichotomy, where isolated sources show higher D/H ratios than clustered counterparts. Establishing which physical and chemical processes create this differentiation can provide new insights into the chemical evolution of water during star formation and the chemical diversity during the star formation process and in young planetary systems.}
   {We seek to determine to what degree the local cloud environment influences the D/H ratio of water in the hot corinos toward low-mass protostars and establish which physical and chemical conditions can reproduce the observed HDO/H$_2$O and D$_2$O/HDO ratios in hot corinos.}
   {The evolution of water during star formation is modeled using 3D physicochemical models of a dynamic star-forming environment. The physical evolution during the protostellar collapse is described by tracer particles from a 3D MHD simulation of a molecular cloud region. Each particle trajectory is post-processed using {\sc radmc-3d} to calculate the temperature and radiation field. The chemical evolution is simulated using a three-phase grain-surface chemistry model and the results are compared with interferometric observations of H$_2$O, HDO, and D$_2$O in hot corinos toward low-mass protostars.}
   {The physicochemical model reproduces the observed HDO/H$_2$O and D$_2$O/HDO ratios in hot corinos, but shows no correlation with cloud environment when similar initial conditions are tested. The observed dichotomy in water D/H ratios requires variation in the initial conditions, for example the duration and temperature of the prestellar phase. Reproducing the observed D/H ratios in hot corinos requires a prestellar phase duration $t\sim 1-3$~Myr and temperatures in the range $T \sim 10-20$~K prior to collapse. Furthermore, high cosmic-ray ionization rates ($\xi_{\mathrm{H}2} \sim 10^{-15}~\mathrm{s}^{-1}$) appear to be incompatible with the observed D/H ratios toward low-mass protostars.}
   {This work demonstrates that the observed differentiation between clustered and isolated protostars stems from differences in the molecular cloud or prestellar core conditions and does not arise during the protostellar collapse itself. The observed D/H ratios for water in hot corinos are consistent with chemical inheritance of water, and no resetting during the protostellar collapse,  providing a direct link between the prestellar chemistry and the hot corino.}


   \maketitle
%
\section{Introduction}

Understanding the formation and evolution of water during star and planet formation is essential to our understanding of the conditions for life in other planetary systems. Water is a prerequisite for life as we know it and, furthermore, also an important molecule for the planet formation process: It contributes significantly to the solid mass reservoirs outside the ice line and impacts the thermal evolution of the gas and the coagulation of dust particles \citep[see, e.g.,][]{dishoeck2014}.

It is well established that water predominantly forms on dust grain surfaces during the molecular cloud phase, where water constitutes the bulk of the ice \citep[e.g.,][]{dishoeck2014}. The evolution from the onset of star formation, through the protostellar collapse and protoplanetary disk phases, and finally onto planetary bodies is however still uncertain. Key questions include to what degree water is processed during star and planet formation, namely, whether planets accrete pristine water inherited from the molecular cloud, or if the water is a product of local processes within the envelope and disk. Crucially, it remains unclear to what extent variations in the local cloud environment impact the water chemistry of the final planetary system, and hence ultimately influence the conditions for biology in extrasolar systems.

Water deuterium fractionation serves as a powerful tracer of the chemical and physical evolution of water during the star and planet formation process, since the degree of deuterium fractionation depends sensitively on the formation environment \citep[e.g.,][]{caselli2012}. Water formed under molecular cloud conditions, where the mean temperature is $T_\mathrm{gas}$~$\sim20$~K and the visual extinction is low ($\sim$ 1--3 mag), shows a moderate degree of deuterium fractionation, with D/H ratios around $10^{-4}$--$10^{-3}$, compared to the canonical value of $1.5\times10^{-5}$ in the local interstellar medium (ISM) \citep{linsky2003}\footnote{As either hydrogen atom in H$_2$O can be replaced by a deuterium atom, the canonical HDO/H$_2$O ratio is 2$\times$(D/H)$_\mathrm{ism}=3\times10^{-5}$}. Conversely, water formed in prestellar cores, where temperatures are lower (${\sim}$~10~$\mathrm{K}$) and the visual extinction higher (Av > 5~mag), is highly enriched in deuterium, with D/H ratios around $10^{-2}$--$10^{-1}$. Hence, the D/H ratio of water can record information about the formation environment. 
Deuterium fractionation during star formation is primarily driven by the gas-phase exchange reaction H$_{3}^{+}$ + HD $\rightleftharpoons$ H$_2$D$^{+}$ + H$_2$ + $\Delta E$, where the exact value of $\Delta E$ depends on the spin state of the involved reactants \citep{pagani1992, hugo2009}. This reaction is exothermic and the backward reaction is inhibited at low temperatures ($T {\lesssim} 50$~K), leading to an enrichment of H$_2$D$^{+}$ relative to H$_{3}^{+}$. H$_2$D$^+$ dissociatively recombines with free electrons to form atomic D, thus increasing the local atomic D/H ratio in the gas phase and ultimately on dust grain surfaces where water and other molecules are formed through hydrogenation. Detections of gas-phase H$_2$D$^{+}$ confirm this enrichment occurs in prestellar cores \citep{caselli2003, vastel2004}.

\begin{figure*}[ht]
\centering
\resizebox{0.88\hsize}{!}
        {\includegraphics{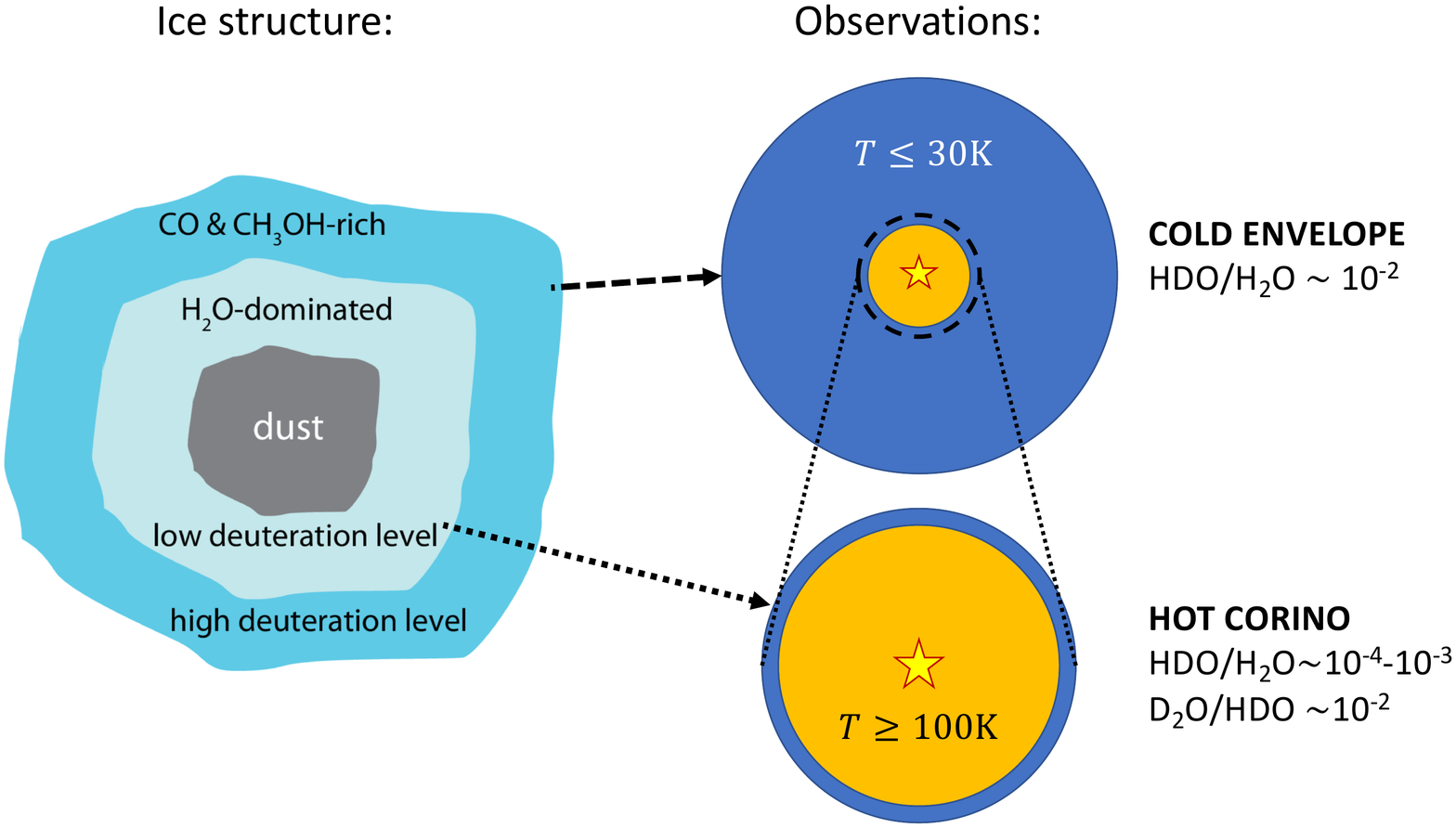}}
  \caption{Schematics of the predicted ice structure in the prestellar core stage (right) and the observed water deuterium fractionation toward embedded low-mass protostars (left). The outer layer of the ice mantle, which has high water D/H ratio, could sublimate in the cold envelope, while the whole mantle is sublimated in the hot corino. The schematical structure is proposed by \citet{furuya2016}
  . Observational values are from \citet{persson2014, coutens2013, coutens2014, jensen2019}.}
     \label{fig:schematic}
\end{figure*}

Observations toward young embedded protostars have revealed a complex evolution of water deuteration during the earliest stages of star formation. 
Single-dish observations of the cold extended envelope show HDO/H$_2$O ratios of the order of $10^{-2}$ \citep{parise2005}, while interferometric observations of the hot corino regions show lower ratios in the range $10^{-4}-10^{-3}$ \citep{coutens2013, persson2014, jensen2019}. Meanwhile, the gaseous D$_2$O/HDO ratio in hot horinos appear to be of the order of $10^{-2}$ which is an order of magnitude higher than the HDO/H$_2$O ratio in the same region  \citep{coutens2014}. This is schematized in Fig. \ref{fig:schematic}. A priori, a higher D$_2$O/HDO ratio than HDO/H$_2$O is unexpected: if H$_2$O, HDO, and D$_2$O were formed at the same time through surface reactions the  [D$_2$O/HDO]/[HDO/H$_2$O] ratio would be the statistical value of 0.25 \citep{rodgers2002}.

The observed fractionation ratios have been explained by detailed chemical models invoking a layered ice structure with at least two notable components: a water-rich ice mantle with low D/H ratio and a surface ice component enriched in  deuterium \citep[e.g.,][and Fig. \ref{fig:schematic}]{taquet2013model, furuya2016}. In the cold outer envelope of protostars, the gas-phase HDO/H$_2$O ratio reflects the D/H ratio of the surface ice chemistry, namely, water formed during the prestellar core phase with efficient deuterium fractionation, and direct gas phase water formation in this region. Conversely, the HDO/H$_2$O and D$_2$O/HDO ratios in hot corinos reflect the bulk ice reservoir, since the entire ice mantle is thermally desorbed in this region. The complete desorption of the ice layers lowers the HDO/H$_2$O ratio in the gas-phase, since the bulk water ice formed during the molecular cloud phase with a lower degree of deuterium fraction. Meanwhile, the D$_2$O/HDO ratio remains roughly constant, as both D$_2$O and HDO primarily formed during similar physical conditions, in the prestellar core phase.
This scenario can explain the observed D/H ratios toward young embedded protostars and indicates that the water chemistry observed in hot corinos predominantly originates from the molecular cloud, in other words, an inheritance scenario in which the water is largely inherited from the molecular cloud. Physicochemical models of protoplanetary disks generally support this notion: the D/H ratios measured in comets cannot alone be explained by local processing within the disk, and hence some degree of inheritance from the prestellar phase is required to reproduce the high D/H ratios of water observed in pristine Solar System material \citep[see, e.g.,][]{cleeves2014, furuya2017, altwegg2017}.

As the water chemistry appears to be mainly inherited from the environment (i.e., the molecular cloud), the question becomes whether and how the large-scale environment can impact the water chemistry of the final planetary system. In a recent work, evidence for a correlation between the HDO/H$_2$O ratios and the local cloud environment is observed.  \citet{jensen2019} find a higher HDO/H$_2$O ratio toward isolated protostars than what has previously been detected toward sources in clustered environments.
The authors proposed two explanations for such a correlation: 1) either temporal differences between isolated and clustered cores: a slower collapse of an isolated prestellar core could prolong the prestellar core phase leading to a higher D/H ratio in the water, or 2) higher temperatures or a stronger radiation field in clustered cores, where nearby (proto)stars and turbulent cloud dynamics may heat the gas compared to isolated counterparts. Chemical diversity induced by variations in the natal cloud environment have scarcely been studied theoretically, as the majority of physicochemical models of protostellar collapse have utilized one-dimensional models that fail to capture the complex nature of star formation in dynamic molecular clouds. 
If local cloud variations drive a difference in the D/H ratio of water in the hot corino phase, such a difference may also impact the complex organic molecules (COMs) which are characteristic of hot corinos and hot cores. In a recent study, \citet{aikawa2020} simulate the impact of various prestellar core conditions on the abundance of COMs and warm carbon chain chemistry (WCCC) in hot corinos. These authors find that the WCCC is more pronounced in cores with lower initial temperature, lower extinction, or a longer prestellar core phase. No clear pattern emerged for the hot corino chemistry; some molecules, such as CH$_3$OH, are less abundant when the temperature is higher, while other molecules show no impact of variation in prestellar core conditions.

In this work, we combine a detailed 3D MHD model of a star-forming region with radiative transfer and a three-phase chemical model to explore how the local environment may change the chemical evolution through protostellar collapse and determine which conditions can reproduce the observed variation in HDO/H$_2$O ratios toward low-mass protostars.
Throughout the paper, we denote HDO/H$_2$O as $f_{\mathrm{D}1}$ and D$_2$O/HDO as $f_{\mathrm{D}2}$. Finally, we define the ratio $f_{\mathrm{D}2}$/$f_{\mathrm{D}1}$ as $\alpha$.

\section{Methods} \label{sec:2}
\subsection{Physical collapse models}
\subsubsection{Molecular cloud models}
\begin{figure*}[ht]
\resizebox{\hsize}{!}
        {\includegraphics{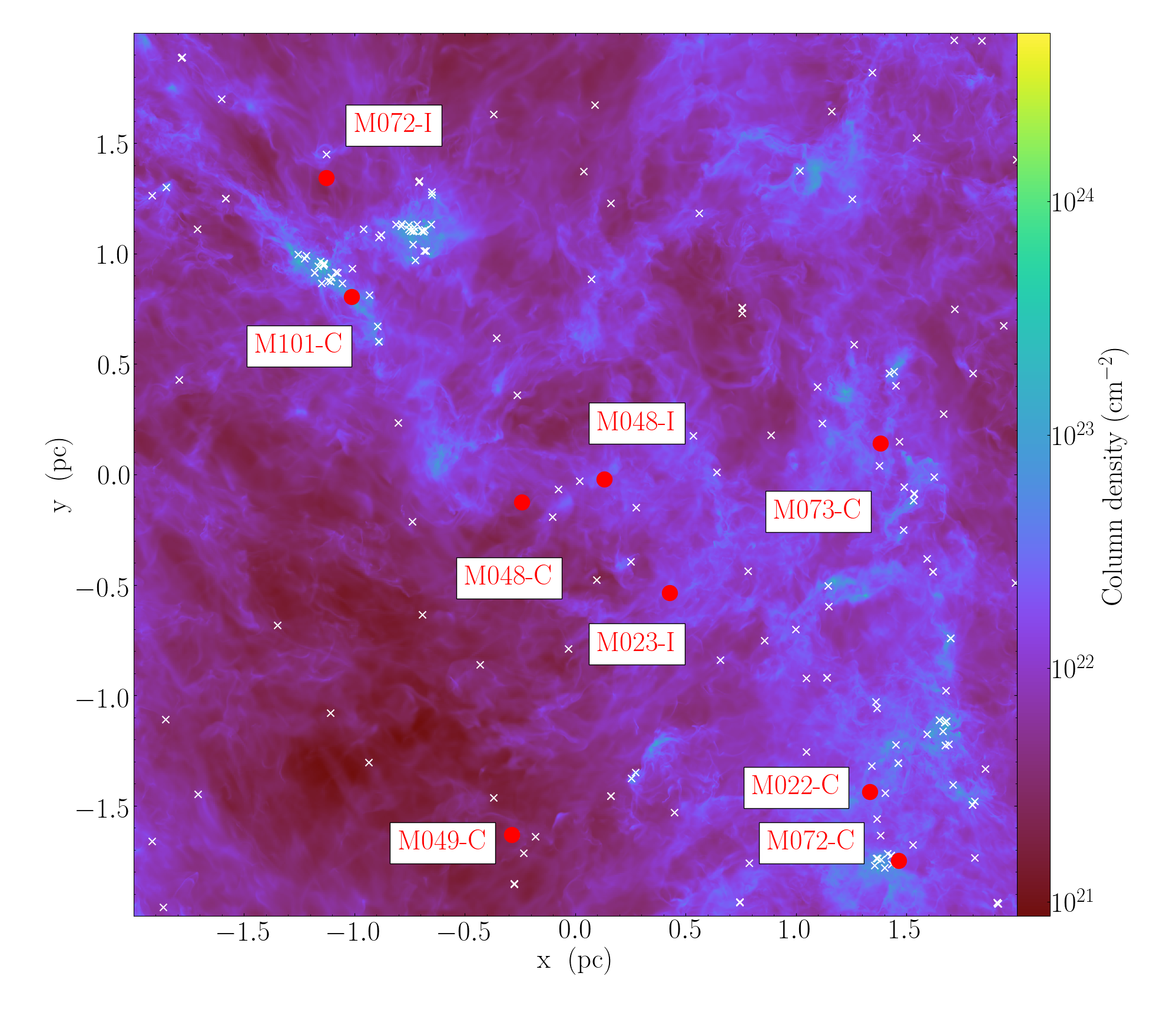}}
  \caption{Projected densities in the molecular cloud simulation. A total of 233 protostars, denoted with white crosses, form by the end of the simulation (t = 1.65 Myr). Most protostars form in clusters along filaments. The protostars studied in this work are denoted with red circles.}
     \label{fig:ramses}
\end{figure*}

\begin{figure}[ht]
\resizebox{\hsize}{!}
        {\includegraphics{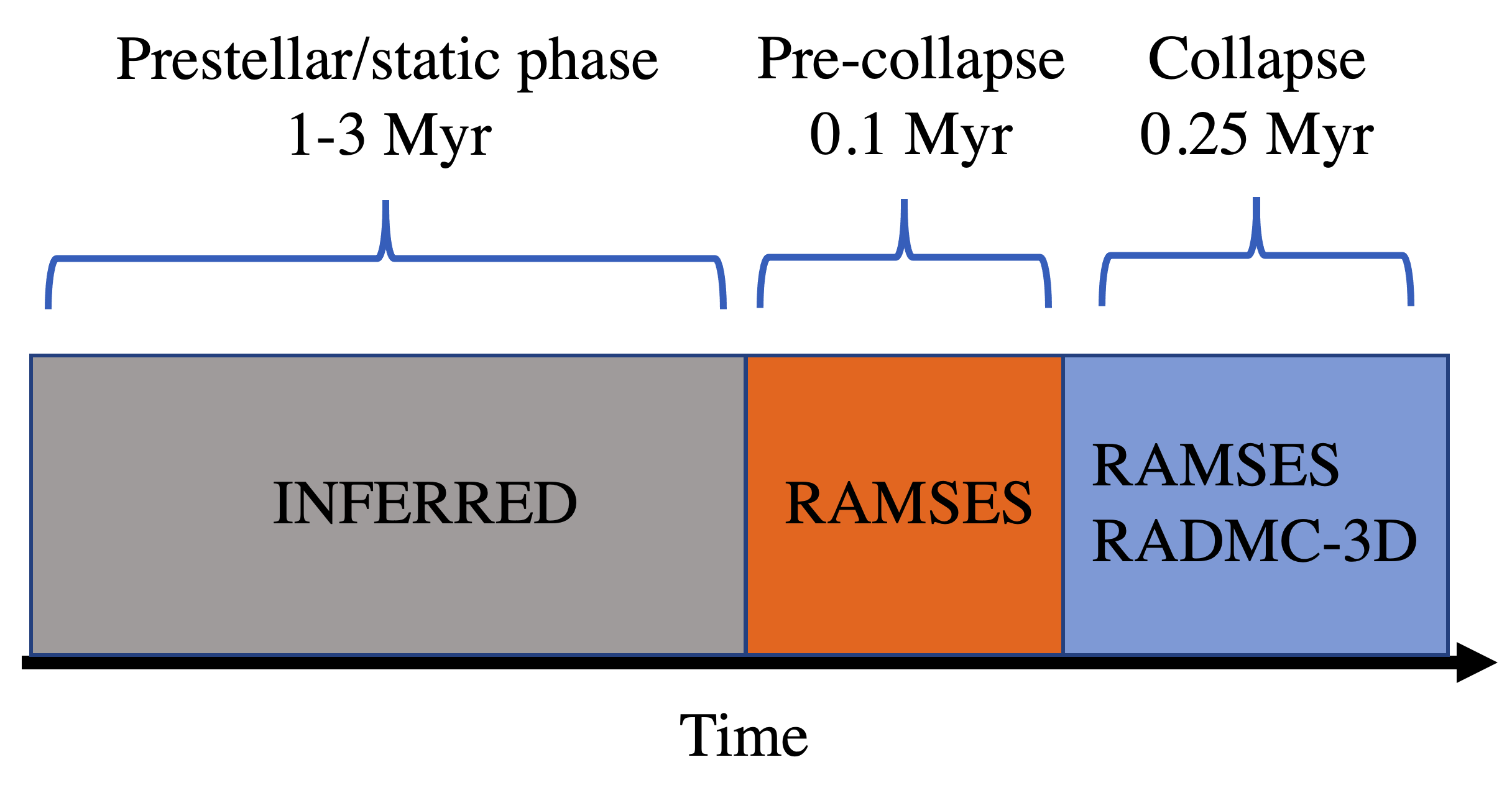}}
  \caption{Overview of the different phases of the model. In the initial prestellar (static) phase, the physical conditions are inferred previous simulations and observations \citep[e.g.,][]{keto2008}. In the pre-collapse phase, which has a fixed duration of 100~kyr, the physical evolution is derived from {\sc ramses}, but since the protostar is not yet formed, no radiative transfer is performed. During the collapse, the physical evolution is derived from {\sc ramses} and post-processed using {\sc radmc-3d}. The duration of the collapse varies between individual tracer particles, with an upper duration of $\sim 250$~kyr.}
     \label{fig:model_overview}
\end{figure}

A simulation of a 4 pc$^3$ region of a molecular cloud is carried out with the adaptive mesh refinement (AMR) code {\sc ramses} \citep{teyssier2002}. The entire box is shown in Fig. \ref{fig:ramses} toward the end of the simulation, where 233 protostars formed in the box. Initial conditions for the molecular cloud, including turbulence driving the star formation process, is identical to the models presented in \citet{haugbolle2018}. The {\sc ramses} model presented in this work was modified to include sink particles and tracer particles. Sink particles represent protostars in the model and record the mass accretion onto the protostars.
Tracer particles are passive particles injected into the simulation where they are advected along the gas flow, recording the dynamical evolution for chemical post-processing. A detailed description of the model, including these modifications, is available in \citet{haugbolle2018}.
The molecular cloud simulation reproduces a number of significant empirical relations for star-forming clouds, such as the Larson scaling relations, the star formation rate (SFR), the core mass function (CMF) and the initial mass function (IMF). The model is therefore well suited to study star formation as a heterogeneous process, where the physical evolution on larger scales may impact the physical and chemical evolution on smaller scales.
Compared to \citet{haugbolle2018}, the physical model used in this work has a factor of two higher resolution with a root grid of 512$^3$. With six levels of refinement this results in a minimum cell-size of 25 au. This resolution is sufficient to study the dynamics from the larger molecular cloud scales down to the collapse of individual protostellar cores, but is not suitable to resolve any emerging accretion disk.  
The accretion rate onto sink particles within the simulation was recorded at a cadence of $\sim$100 years, while the physical structure of the simulation was recorded in `snapshots' at a lower cadence, once every 5 kyr. We calculated the positions of the tracer particles between each snapshot from the position, velocity, and acceleration vectors at each snapshot. For each sink particle, we followed tracer particles for a duration of $\sim$350 kyr, starting with a pre-collapse phase 100 kyr before the formation of the sink particle (i.e., the onset of protostellar collapse) and continuing through the Class 0 and Class I phases. Including a longer pre-collapse phase was not possible, as several of the studied sink particles form early in the simulation. An overview of the different phases of the simulation is shown in Fig. \ref{fig:model_overview}. During the pre-collapse phase, we averaged the density in 10~kyr windows to reduce computational time in this phase, where the temperature remains fixed.

As no disks form within the simulation, the evolution toward the end of the simulated time range is less accurate as the formation of an accretion disk changes the dynamics and radiation field in the inner part of the core. We note that higher-resolution simulations performed within the same framework, such as the models presented in \citet{kuffmeier2018}, show that accretion disks form when the resolution is increased. We focus on the earlier stages of star formation in this work, namely the protostellar collapse down to hot corino scales ($\sim100$~au).
A sample of low-mass protostars, both clustered and isolated, were selected from the simulation. Throughout this work, we define isolated protostars as sink particles where no other protostars enter within 20,000 au during the simulation. For the clustered sources, we require at least two neighboring protostars within 10,000 au at some point in the simulated evolution. The characteristics of the selected sample is summarized in Table \ref{tab:sinks}

\begin{table}
\caption{Overview of the simulated protostars studied in this work. Isolated and clustered sources are denoted I and C, respectively. The envelope mass, $M_\mathrm{env}$, is the total mass within a 10,000 au sphere at the onset of collapse (i.e., at the moment the sink particle is formed). The free-fall timescale is calculated assuming a uniform density within 10,000~au. The quantity $M_\mathrm{final}$ indicates the final mass at the end of the simulation.}\label{tab:sinks}
\centering
\begin{tabular}{lllll}
\hline\hline
Sink & $M_\mathrm{final}$($\mathrm{M}_\odot$) & $M_\mathrm{env}$ ($\mathrm{M}_\odot$) & $t_\mathrm{ff}$ (kyr) & I/C \\ \hline
M022--C   & 0.22                  & 1.50                      & 144.5        & C        \\
M023--I    & 0.23                  & 0.90                      & 186.4        & I        \\
M048--C   & 0.48                  & 1.06                      & 171.6        & C        \\
M048--I   & 0.48                  & 1.19                      & 162.2        & I       \\
M049--C   & 0.49                  & 0.73                      & 206.5        & C        \\
M072--I    & 0.72                  & 0.85                      & 192.2        & I        \\
M072--C   & 0.72                  & 5.58                      & 74.8         & C        \\
M073--C  & 0.73                  & 0.94                      & 182.5        & C        \\
M101--C   & 1.01                  & 1.74                      & 134.1        & C        \\
\hline
\end{tabular}
\end{table}
\subsubsection{Radiative transfer models}

\begin{figure*}[ht]
\resizebox{\hsize}{!}
        {\includegraphics{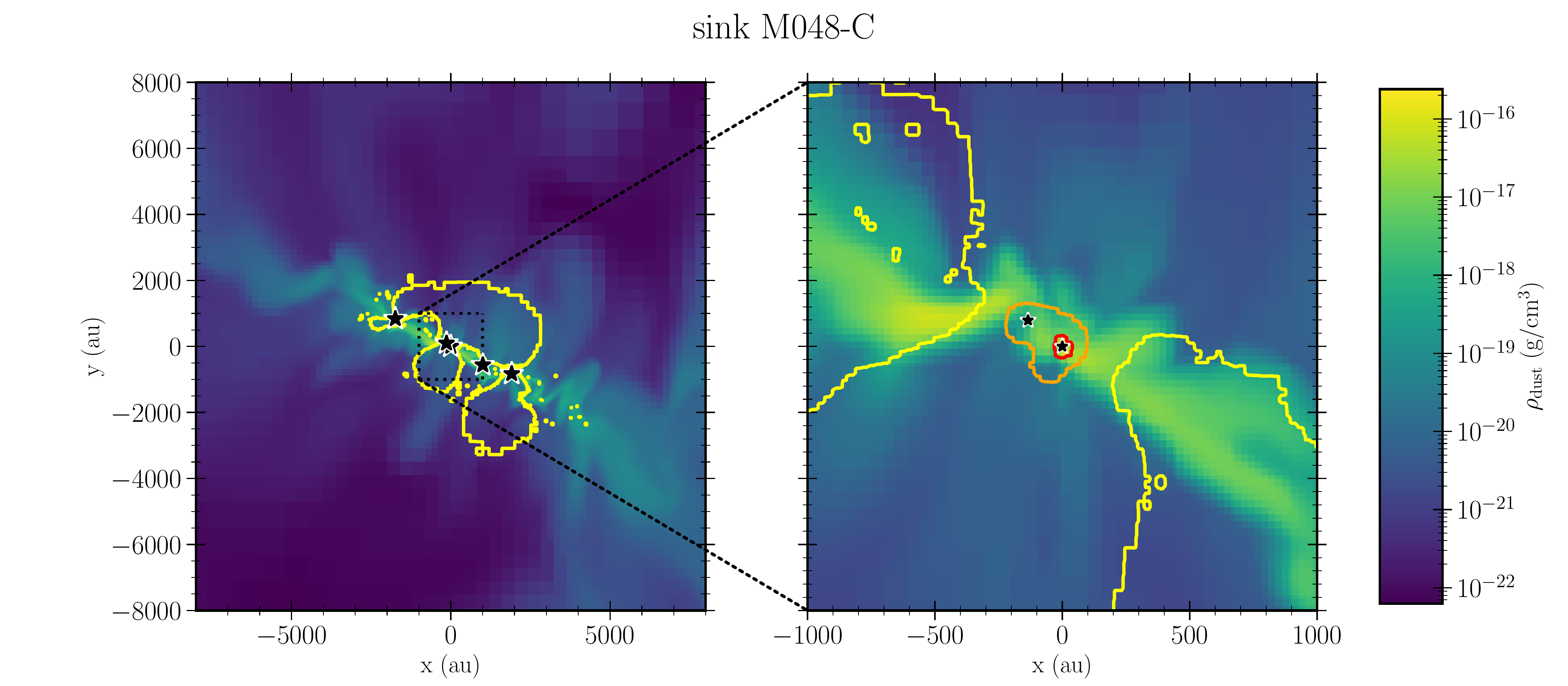}}
  \caption{Structure around the binary sink M048-C with a final mass of M$ = 0.48\mathrm{M}_\odot$. The color map shows the density and reveals that the sink, along with several siblings, is born along a filament, extending from northwest to southeast in the imaged plane. Contours show the temperature, calculated via {\sc radmc-3d}, in levels of [25.0, 50.0, 100.0]~K. In the right panel, the yellow 25~K contour is incomplete; the left panel shows the full structure of this region. The image shows a slice through the grid, not a projection.}
     \label{fig:radmc_model}
\end{figure*}   

The dust temperatures and radiation field around each sink particle was post-processed using the Monte Carlo radiative transfer code {\sc radmc-3d} \citep{dullemond2012}. 
For each sink particle in the {\sc ramses} simulation, a spherical cutout of 30,000 au around the sink particle was converted to a {\sc radmc-3d} AMR grid. Creating smaller cutouts of the total grid is necessary to reduce the memory load, which can be substantial for large photon packages. The dust temperature is calculated via the {\sc mctherm} routine; we injected 15 million photons in each radiative transfer model to sample the grid adequately. Models with a higher number of photons were tested and the results of the radiative transfer converge to the same result. We utilized the dust opacities from \citet{semenov2003} in the radiative transfer modeling.

The protostars, represented by the sink particles, are included in the radiative transfer models as point sources. In addition to the central protostar in each cutout, we included neighboring protostars entering within 10,000~au of the protostar in focus. The protostellar luminosity is interpolated from the pre-main sequence (PMS) models from \citet{pms1997}. We included a 100 kyr offset between the onset of core collapse and the PMS age, similar to \citet{young2005}, corresponding to approximately one free-fall timescale. In the first 100 kyr of the collapse, the surface temperature of the protostar is fixed at 2000~K. 
The radius is set to be 2.5 $R_\odot$ for the calculation of the accretion luminosity. This implementation is similar to that presented in \citet{frimann2016}.

In addition to the protostellar luminosity derived from the PMS models, we included the accretion luminosity of the central protostar from Eq. \ref{eq:lacc},
\begin{equation}\label{eq:lacc}
    L_\mathrm{acc} = G \frac{M \dot{M}}{2R}.
\end{equation}
The accretion rates onto the sink particles were recorded with a higher cadence than the {\sc ramses} output. To account for this difference in sampling, we ran a total of five {\sc radmc-3d} models for each {\sc ramses} output with varying accretion luminosities: we created a linear grid of four models, ranging from the lowest accretion rate to the highest accretion rate associated with any given {\sc ramses} output. We calculated the fifth model using the mean accretion rate during the period between the snapshots. When estimating the dust temperature for a tracer particle at a given position and time, we calculated the total luminosity ($L_\mathrm{acc}$ + $L_\mathrm{protostar}$) of the protostar at that time, and interpolated the temperature from the values provided by the five {\sc radmc-3d} models. We assume that the dust and gas is thermally coupled, such that the gas temperatures and dust temperatures are equal. This approach is reasonable at higher densities present during the protostellar collapse \citep[$n \gtrsim 10^{4}$,][]{goldsmith2001}, but may not be valid in the prestellar phase where densities are lower. The limitations of the model are discussed in Sect. \ref{sec:4_2}.
A lower limit on the temperature of 10~K is set throughout the chemical modeling. While lower temperatures can occur in the center of dense cores, this is a reasonably average temperature for dense cores \citep[e.g.,][]{pagani2007}. Furthermore, many reactions in the chemical network are validated to a lower limit of 10~K.

Along with the thermal dust calculation, the mean UV radiation field is calculated in the {\sc radmc-3d} grid with the {\sc mcmono} routine. In addition to the protostellar sources, the models include an external radiation field in the form of the interstellar radiation field (ISRF) from \citet{draine1978}, denoted as $G_0$, which irradiates the 30,000~au {\sc radmc-3d} domain around each sink particle.
For the {\sc mcmono} calculations, we opted to run just one model using the mean accretion luminosity for each {\sc ramses} snapshot. As the UV flux emerging from the central protostar is heavily attenuated, the calculated UV field is dominated by the external radiation field, except for the innermost 25-50 au around a protostar. Running five models with varying accretion luminosity would hence not have a notable impact on the surrounding envelope and the chemical evolution herein. Figure \ref{fig:radmc_model} shows an example of the thermal structure calculated around one of the sinks studied in this work.

We note that the earliest protostellar evolution may not be well explained by the PMS models, as highlighted by the luminosity problem \citep[see, e.g., ][and references therein]{jensen2018}. We do not consider this a major issue for this work for two reason. First, the accretion luminosity, not the protostellar luminosity, is the primary luminosity source at the very early stages modeled here. Second, the purpose of this work is to study how variations in environment influences the chemistry during star formation and does not depend on the exact choice of PMS model. Additionally, we note that the PMS phase occurs toward the end of the range of the time-domain studied in this work, namely, the $\sim$250~kyr of the protostellar collapse after the initial collapse of the core.

\subsection{Calculating extinction in the cores}
For photodissociation and photodesorption, rate coefficients are given as a function of the visual extinction $A_\mathrm{v}$. From the calculated mean field intensities, we follow the approach of \citet{visser2011} and derive the optical depth $\tau_\mathrm{UV}$ at UV wavelengths from:
\begin{equation}
\label{eq:tau}
	\tau_{\mathrm{UV}} = - \ln{\frac{F_\mathrm{UV,RADMC-3D}}{ F_\mathrm{UV,0}}} \, ,
\end{equation}
where $F_\mathrm{UV,RADMC-3D}$ is the integrated UV flux from the radiative transfer calculation, and $F_\mathrm{UV,0}$ is either the integrated UV flux from the ISRF or from the central protostar, whichever is higher in each grid point. Initially, the UV flux from the protostar is quite low owing to the low surface temperature of the protostar (2000 K) and during this period, the ISRF dominates the UV flux. Later on, the UV flux from the protostar may increase, however, the flux is not able to penetrate the envelope, because no outflow cavities are resolved in the simulation.

From the optical depth, the visual extinction is calculated as $A_\mathrm{v}^\mathrm{RADMC-3D} = \tau_\mathrm{UV} / 3.02$. In addition to the extinction calculated with {\sc radmc-3d}, we assume an ambient extinction everywhere in the molecular cloud, (i.e., surrounding the 30,000 au radiative transfer domain). The total extinction during the collapse is then given as $A_\mathrm{v} = A_\mathrm{v}^\mathrm{cloud} + A_\mathrm{v}^\mathrm{RADMC-3D}$, where we assume $A_\mathrm{v}^\mathrm{cloud} = 5$~mag in our fiducial model. Models with a lower ambient extinction, $A_\mathrm{v}^\mathrm{cloud} = 2$~mag, are presented in appendix \ref{app:av}.
The column densities are estimated using the empirical relation $N_\mathrm{H} = A_\mathrm{V}\times 1.59\times10^{21}$~$\mathrm{cm}^{-2}$ \citep{bohlin1978, diplas1994}.
These parameters are necessary to derive the self-shielding functions and photochemical rates for the chemical models.

As mentioned in the previous section, each tracer particle trajectory includes 100~kyr of evolution prior to the formation of the sink particle. Since the protostar is not formed in the pre-collapse phase, no radiative transfer is calculated, and we use a fixed temperature of 10~K and a fixed visual extinction of $A_\mathrm{v} = 5$~mag during this period.

\subsection{Chemical model}
The chemical evolution was simulated using the rate-equation approach \citep[see, e.g.,][]{cuppen2017}. The model adopts the three-phase model introduced by \citet{hasegawa1993b}, with a number of modifications detailed below. 
Three-phase models consider the ice mantle and ice surface as distinct phases, as opposed to two-phase models where the ice species all reside in the same (bulk) phase \citep{hasegawa1992}. As the chemistry evolves, two-phase models may produce inaccurate results, since these models cannot distinguish between species present in the reactive surface layers and species buried deep in the ice mantle where the molecules are inert or less likely to react. Three-phase models overcome some of these issues by separating ice species in surface and mantle phases, however other caveats are introduced with these models, which we will discuss later.
We do not consider bulk diffusion in our model; the effectiveness of bulk reactivity is still a matter of debate \cite[see, e.g.,][]{shingledecker2019}.

Experiments have shown that surface chemistry likely occurs in the top two to four monolayers\footnote{One monolayer is defined as the amount of binding sites on the grain surface, such that one monolayer corresponds to every site on the dust grain being occupied}. Motivated by these results, we adopted a similar approach to that of \citet{furuya2015}, in which we extended the surface layer to thickness of four monolayers. Other three-phase models , such as the {\sc nautilus}, model have adopted a similar approach, albeit with two monolayers instead of four \citep{ruaud2016}. Henceforth, we refer to the surface phase of the model as the surface layer, although the thickness of the phase is four monolayers.

\subsubsection{Chemical network}
For the gas-phase chemistry, we adopted the chemical network presented by \citet{majumdar2017}, which extends the {\sc kida}\footnote{\url{http://kida.astrophy.u-bordeaux.fr}} network to include spin-state chemistry of H$_2$, H$_{3}^{+}$ and other key molecules necessary to accurately model deuterium fractionation processes in the cold ISM. The basis of the spin-state chemistry was presented in \citet{sipila2015}. The gas-phase network was initially benchmarked in \citet{majumdar2017} for dark cloud conditions and found to agree with observations. 
In this work, the network was updated to include more recent photodissociation rates from the Leiden photodissociation database \citep{heays2017}, a different treatment of grain recombination reaction, and other minor adjustments. We refer to the papers of \citet{wakelam2015} and \citet{majumdar2017} for a detailed description of the network and reactions. We excluded three-body reactions from these calculations, as they are inefficient under typical ISM conditions \citep{cuppen2017}.

 For the surface network, we used a network derived from \citet{taquet2013model}, which was further extended to include reactions from \citet{garrod2013}. The network has been extended to include single and doubly deuterated isotopologues for molecules with up to four carbon atoms. For a number of key species such as methanol, we included all deuterated isotopologues, such as CD$_3$OD. 
 
Binding energies from \citet{garrod2006} were used for the majority of the species, with updated estimates from \citet{wakelam2017} used where applicable. 
For deuterated isotopologs, we used the same binding energy as the main species, except for atomic D, which is set 21~K higher than H atoms \citep{caselli2002}. For the extensive network used in this work, a few species do not have any binding energy estimates. For these, we use the binding energy of H$_2$O as a ``default'' value. 

The full chemical network and the chemical code is publicly available for reference and provided for the community for future work \footnote{\url{https://github.com/ssjensen92/kemimo}}.



\subsubsection{Adsorption and desorption}
For the adsorption onto the grain surface and thermal desorption of surface species, we followed the method of \citet{hasegawa1992}.
The adsorption rate coefficient is given by:
\begin{equation}
		k_\mathrm{ads}(i) = S \langle v(i)\rangle\sigma_{\mathrm{dust}} n_{\mathrm{dust}} \quad(\mathrm{s}^{-1}) \, ,
\end{equation}
where $\sigma_{\mathrm{dust}}=\pi r_{\mathrm{dust}}^2$ is the cross-section of dust grains, $n_{\mathrm{dust}}$ is the number density of dust grains, and $\langle v(i)\rangle=\sqrt{8k_\mathrm{b}T/\pi m_{i}}$ is the mean thermal velocity of the species $i$. There parameter $S$ is the sticking coefficient, which is estimated from the experimentally derived prescription presented by \citet{he2016}.

The thermal desorption is calculated from the first-order expression as follows:
\begin{equation}
	k_\mathrm{des}(i) = \nu_i \exp \left(\frac{-E_\mathrm{bind}}{k_\mathrm{b}T}\right) \quad(\mathrm{s}^{-1}) \, ,
\end{equation}
where $E_\mathrm{bind}$ is the binding energy of the species on the surface. $\nu_i = \sqrt{(2 N_\mathrm{s} E_\mathrm{bind})/(\pi^2 m_i)}$ is the characteristic frequency for the species, calculated assuming a harmonic potential for the surface binding sites \citep{tielens1987}. The quantity $N_\mathrm{s}$ is the number of binding sites per dust grain and $m_i$ is the mass of the species.

Over time, incident cosmic-ray particles heat dust grains \citep[e.g.,][]{hasegawa1993b}. This leads to a cosmic-ray induced desorption rate, which we calculated following \citet{leger1985}:
\begin{equation}
	k_\mathrm{des,CR-induced} = f(70 \mathrm{K}) k_\mathrm{des}(70 \mathrm{K}) \quad(\mathrm{s}^{-1}) \, ,
\end{equation}
where the factor $f$ determines the fraction of time the dust grains spend at 70~K. This value is determined through the balance between heating rate of dust grains by cosmic-ray particles and the cooling rate through radiation as follows: $f (70 K) = \frac{\zeta_CR}{1.3\times10^{-17}} \times3.16\times10^{-19}$.

\subsubsection{Surface reactions}
We modeled surface chemistry reactions using the Langmuir-Hinshelwood mechanism, which assumes that both reactants adsorb onto the grain surface before any reaction occurs \citep{cuppen2017}. After adsorption, the reactants diffuse thermally across the grain surface and upon encounter they may react, depending on the reaction probability. The rate coefficient $k_{ij}$ for this mechanism is given as
\begin{equation}
\label{eq:LH}
	k_\mathrm{LH}(i,j) = \kappa_{ij} \left(k_{\mathrm{hop}}(i) + k_{\mathrm{hop}}(j) \right) \frac{1}{N_\mathrm{layer} N_\mathrm{s} n_{\mathrm{dust}}} \quad(\mathrm{cm}^3~\mathrm{s}^{-1}),
\end{equation}
where $N_\mathrm{layer} = n_\mathrm{s}/(n_\mathrm{dust} n_\mathrm{s})$ is the current number of surface layers in monolayers. In this equation, $n_\mathrm{s}$ denotes the sum of the number densities of all surface species, $n_\mathrm{dust}$ denotes the number density of dust grains, $\kappa_{ij}$ is the reaction probability, and $k_{\mathrm{hop}}$ is the thermal hopping rate given by $k_{\mathrm{hop}} = \nu_{i} \exp (-E_\mathrm{diff}/k_\mathrm{b}T)$. We assume $E_\mathrm{diff} = 0.5\times E_\mathrm{bind}$ in this work. The reaction probability $\kappa_{ij}$ is unity for barrierless reaction. For reactions with barriers, the barrier is overcome either through quantum tunneling or thermal hopping, whichever is the fastest. We included reaction-diffusion competition as introduced by \citet{garrod2011}. This mechanism considers that two reactants encountering one-another in a binding site may have multiple chances to react before one of them diffuses out of the site, hence increasing the reaction probability of reactions with barriers. 
The inclusion of reaction-diffusion competition is pronounced. For instance, the primary formation pathway of H$_2$O changes from the barrierless reaction H + OH $\rightarrow$ H$_2$O to the reaction H$_2$ + OH $\rightarrow$ H$_2$O + H, which has a barrier of 2100~K, when the mechanism is included \citep{furuya2015}. The reaction probability when assuming reaction-diffusion competition is calculated from:
\begin{equation}
	\kappa_{ij} = \frac{\kappa_{ij}^{0}}{\kappa_{ij}^{0} + k_\mathrm{hop}(i) + k_\mathrm{hop}(j)} \, ,
\end{equation}
where $\kappa_{ij}^{0}$ is the probability to overcome the reaction barrier. The probability of crossing an activation barrier $E_\mathrm{A}$ through thermal hopping is given by $\kappa_{ij}^{0} = \exp (-E_\mathrm{A}/k_\mathrm{b}T)$, while the expression for quantum tunneling depends on the choice of potential barrier. For a rectangular barrier of width $a$, the expression is $\kappa_{ij}^{0} = \exp [-2(a/\hbar )(2\mu E_\mathrm{A})^{1/2}]$. In this expression, $\mu$ denotes the reduced mass of the reactants. We adopted a rectangular barrier with a width of 1 \AA, except for a number of reactions listed in appendix \ref{app:reactions}. 

For exothermic reactions, products may desorp from the grain surface owing to the excess energy released during the reaction. This process is known as chemical desorption and efficiency of this process is poorly constrained, but could play an important role for the interplay between gas-phase and grain-surface chemistry in star-forming regions. We included chemical desorption using the prescription of \citet{garrod2007} for the probability of desorption for reactions with one product, with an efficiency parameter of $a = 0.01$ \citep[see, e.g., ][]{cuppen2017}. Reactions with multiple products are not allowed to chemically desorp. 

\subsubsection{Photochemistry}
The chemical model includes photodissociation and photodesorption of species in the surface layer. Data on photodissociation cross-sections for surface species are sparse and we adopted a similar approach as in \citet{furuya2015} for the surface photochemistry. In this approach, gas-phase photodissociation rates are rescaled for the corresponding surface reaction. The scaling factor is determined from the ratio of the gas-phase to surface rates for water: $k_{ij}^\mathrm{surface} = k_{ij}^\mathrm{gas} \times k_{\mathrm{H_{2}O}}^\mathrm{surface}/k_{\mathrm{H_{2}O}}^\mathrm{gas}$. 
The rate coefficient for the photodissociation is given as:
\begin{equation}
	k_\mathrm{photodiss}(i) = \theta_i P_\mathrm{abs}(i) \sigma_\mathrm{dust}n_\mathrm{dust} F_\mathrm{UV}(i) \times \min (N_\mathrm{layer}, 4) \quad(\mathrm{s}^{-1}) \, ,
\end{equation}
where $\theta_i$ is the surface coverage factor of the species, $\theta_i = n(i)/n_\mathrm{s}$, $F_\mathrm{UV}$ is the UV flux in photons~cm$^{-2}$~s$^{-1}$, and we allowed photodissociation in the top four monolayers, that is, the surface layer in the three-phase model. In our model, it is implicitly assumed that the photo-fragments in the ice mantle recombine immediately. While this may lead to an increase in the abundance of radicals and trigger the formation of COMs, the effect on the water chemistry studied in this work is expected to be limited as a result of the high extinction in these environments.
The photodesorption rates are given by
\begin{equation}
	k_\mathrm{photodes}(i) = Y_i \theta_i \sigma_\mathrm{dust}n_\mathrm{dust} F_\mathrm{UV}(i) \times \min (N_\mathrm{layer}/4, 1) \quad(\mathrm{s}^{-1}) \, ,
\end{equation}
where $Y_i$ is the effective yield per photon for thick ice. For photodesorption, we likewise allowed reactions in up to four monolayers.
Photodesorption yields for CO, CO$_2$, and N$_2$ are derived from experiments \citep{oberg2007, oberg2009, fayolle2011, fayolle2013}. The CO yield for thick ice is 2.7$\times10^{-3}$, while for CO$_2$ the total yield is 10$^{-3}$, with an equal branching ratio between the outcomes CO$^{\mathrm{gas}}$ + O$^{\mathrm{gas}}$ and CO$_{2}^{\mathrm{gas}}$. The desorption yield for N$_2$ is set to 5$\times10^{-4}$.
For H$_2$O, HDO, and D$_2$O, we adopted the branching ratios for photodissociation and photodesorption presented by \citet{arasa2015}. The absorption probability for H$_2$O per incident UV photon is set to $5\times10^{-3}$.
For the remaining species we used a generic yield of 10$^{-3}$. While the network was developed to include complete photochemistry, we note that in the models presented in this work the photon flux is heavily attenuated for the majority of the time, and therefore the effect of photochemistry is limited.
The model included self-shielding for H$_2$, HD, CO, and N$_2$. For CO and N$_2$, we interpolated the shielding factor from the tables of \citet{visser2009b} and \citet{heays2014}. For H$_2$ and HD we used the parametric approximations derived by \citet{draine1996, richings2014} and \citet{wolcott-green2011}, respectively.



\subsection{Benchmarking the network}
\begin{figure*}[ht]
\resizebox{\hsize}{!}
        {\includegraphics{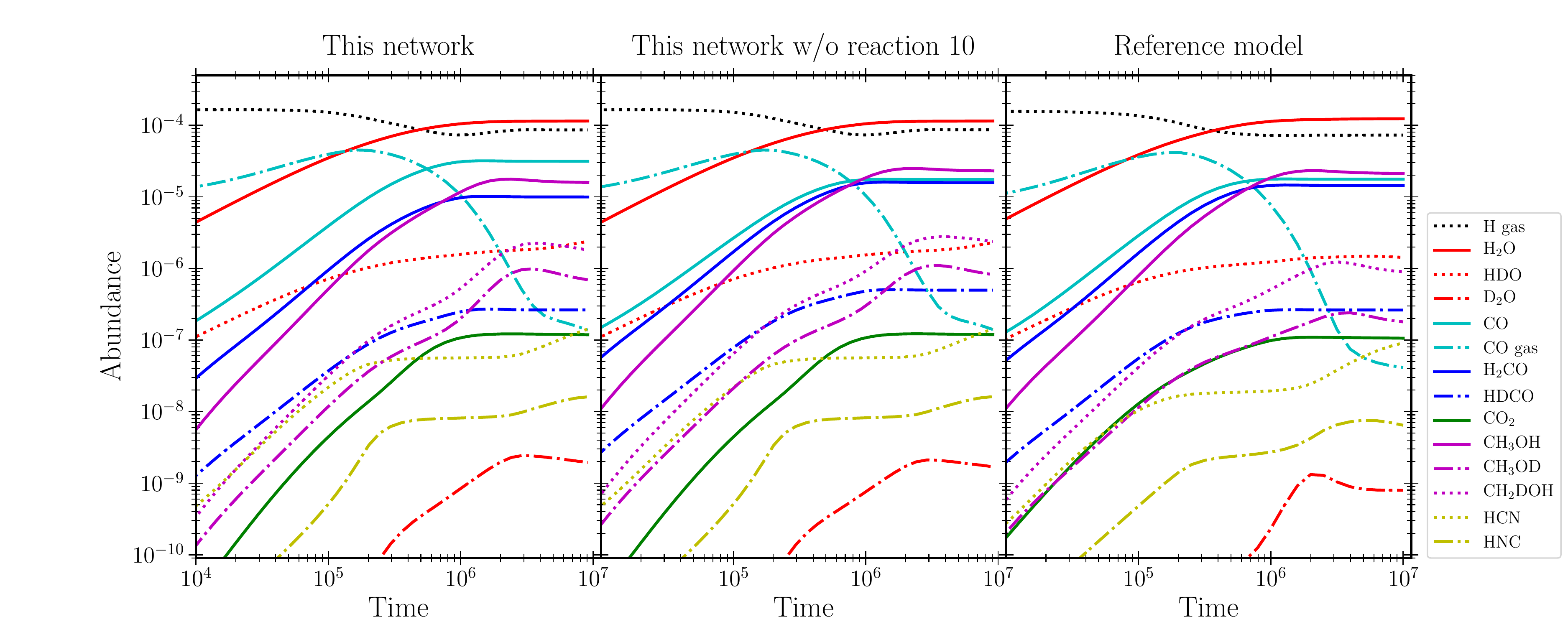}}
  \caption{Steady-state surface chemistry at 10~K, n$_\mathrm{H} = 2\times10^4$, and $A_\mathrm{v} = 5$~mag. All species are bulk (mantle + surface) ice, expect when specified otherwise. The left panel shows the model used here with the inclusion of CO formation through reaction \ref{eq:COreac}, while the middle panels show the network with this reaction switched off. The right panel shows the reference model from \citep{furuya2015, furuya2016}}
     \label{fig:static_comparison}
\end{figure*}

The chemical model was benchmarked against the gas-grain model presented in \citet{furuya2015} and \citet{furuya2016}, which successfully reproduced the observed HDO/H$_2$O and D$_2$O/HDO ratios toward protostellar cores. We refer to this model as the reference model throughout this section. 
Initially, a 10~Myr steady-state simulation with $n_\mathrm{H} = 2\times10^{4} ~\mathrm{cm}^{-3}$, $T = 10$~K, and $A_\mathrm{v} = 5$~mag  was compared, shown in Fig. \ref{fig:static_comparison}. The cosmic-ray ionization rate for $\mathrm{H}_2$ is set to $\xi_{\mathrm{H}_2} = 1.3\times10^{-17} ~\mathrm{s}^{-1}$. The abundances of major species, such as H$_2$O, CH$_3$OH, and CO$_2$ show good agreement, while the CO abundance differ by a factor of $\sim2$. This is due to the inclusion of the hydrogen abstraction reaction:
\begin{equation}\label{eq:COreac}
    \mathrm{HCO} + \mathrm{H} \rightarrow \mathrm{CO} + \mathrm{H}_2 .
\end{equation}
This reaction is not included in the reference model, but was recently demonstrated as a pathway for CO formation \citep{chuang2016}. As in the reference model, H$_2$O and HDO are primarily formed in the surface chemistry, while D$_2$O is dominantly formed in the gas phase. The principal formation pathway of H$_2$O is the surface reaction $\mathrm{H}_2 + \mathrm{OH} \rightarrow \mathrm{H}_2\mathrm{O} + \mathrm{H}$, with a barrier of 2100~K. This pathway dominates when reaction-diffusion competition is included in the model. Without reaction-diffusion competition, the dominant pathway to H$_2$O formation would be the surface reaction $\mathrm{OH} + \mathrm{H} \rightarrow \mathrm{H}_2\mathrm{O}$, as in \citet{furuya2015}. The deuterated variant $\mathrm{H}_2 + \mathrm{OD} \rightarrow \mathrm{HDO} + \mathrm{H}$ can form HDO efficiently, but D$_2$O formation is inefficient owing to the lower tunneling probability for the heavier HD and D$_2$ molecules. Gas-phase formation of water proceeds through ion-molecule reactions, namely, D$_2$O is formed in the reaction HD$_2$O$^+$ + e$^-$ $\xrightarrow[]{}$ D$_2$O + H \citep[see, e.g.,][]{dishoeck2014}. Overall, the reactions leading to the formation of water and deuterated isotopologs are similar to the reference model.

\begin{figure*}[ht]
\resizebox{\hsize}{!}
        {\includegraphics{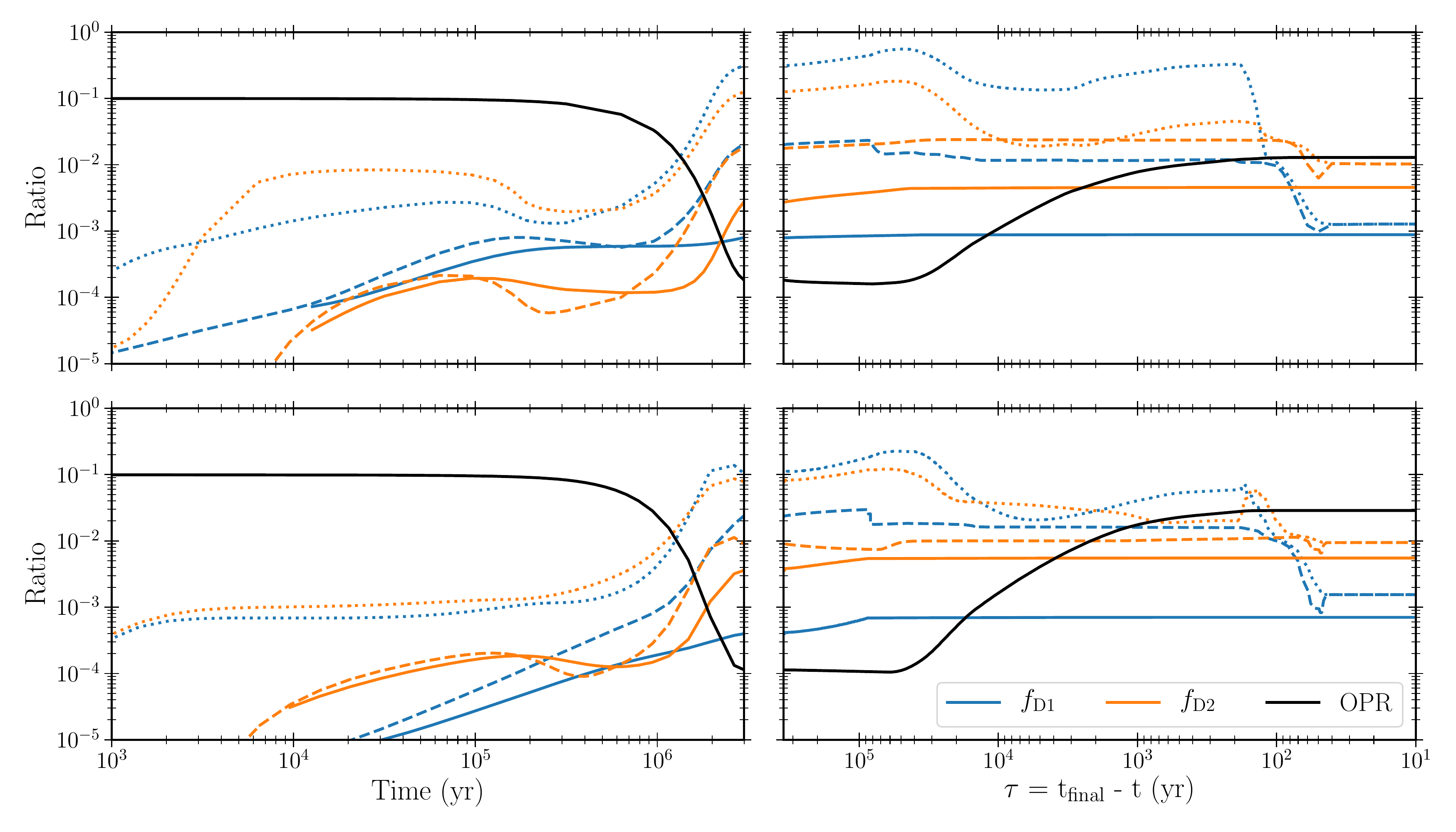}}
  \caption{Comparison between the chemical models presented in \citet{furuya2016} (top panels) and the model utilized here (bottom panels). The 1D RHD simulations of a protostellar collapse are from \citet{masunaga2000}. The solid lines indicate the ice mantle, the dashed lines present the ice surface, and the dotted lines show the gas-phase ratios in the three-phase models. The log-scale on the x-axis is reversed on the second row.}
     \label{fig:1d_comparison}
\end{figure*}

The chemical models were subsequently compared using the 1D radiative hydrodynamics (RHD) model utilized in \citet{furuya2016}. The physical model follows a tracer particle during the collapse phase of a $3.9 \mathrm{M}_\odot$ core that forms a solar mass star; the simulation is explained in detail in \citet{masunaga2000}. The initial conditions prior to the protostellar collapse are a 3~Myr static phase with $n_\mathrm{H} = 2\times10^{4}~\mathrm{cm}^{-3}$, $T = 10$~K, and an visual extinction of $A_\mathrm{v} = 5$~mag. The model is started from the atomic abundances detailed in Table \ref{tab:abundances}. We adopted the ISRF of \citet{draine1978}, corresponding to $F_\mathrm{UV}\sim2\times10^{8}$ photons $\mathrm{cm}^{-2} \mathrm{s}^{-1}$. The cosmic-ray induced UV flux is set to $3 \times 10^{3} $ photons $\mathrm{cm}^{-2} \mathrm{s}^{-1}$. We define these initial conditions as our fiducial model; we return to the impact of the conditions during the static phase in subsequent sections.
Figure \ref{fig:1d_comparison} shows a comparison between the results obtained at the end of the simulation for a particle trajectory that reaches the protostar (4.9 au) after $\sim350$ kyr. The final gas-phase ratios are comparable, with $f_{\mathrm{D}1}$ of $1.3\times10^{-3}$ in the reference model, compared to $1.6\times10^{-3}$ the model used here. For $f_{\mathrm{D}2}$, the results are $10^{-2}$ and $9.5\times10^{-3}$ for the reference model and this model, respectively. The minor difference in ratios are attributed to differences in the gas-phase network, where there is less formation of HDO and D$_2$O compared with the reference model during the final stages of the collapse. 
The model presented here agrees with the chemical evolution of water proposed by \citet{furuya2016}, with lower D/H ratios in the mantle and higher D/H ratios in the surface ice (see Figs. \ref{fig:1d_comparison} and \ref{fig:ice_change}). 

\begin{table}[ht]
\caption{Initial abundances following \citet{aikawa2001}, except for the inclusion of ortho and para spin-states of H$_2$. The initial ortho/para ratio is motivated by the results of \citet{furuya2016}, who found that a ratio of 0.1 best reproduced the observed D/H ratios in hot corinos toward low-mass protostars in their models.}
\begin{tabular}{ll|ll}
\hline
\hline
Species & Abundance (/$n_\mathrm{H}$) & Species & Abundance (/$n_\mathrm{H}$) \\ \hline
o-H$_2$ &      4.5(-1)                  & S+      & 9.14(-8)                    \\ 
p-H$_2$ &      4.5(-2)                  & Mg+     & 1.09(-8)                        \\ 
He      &      9.75(-2)                 & Si+     & 9.74(-9)                        \\ 
HD      &      1.5(-5)                  & Fe+     & 2.74(-9)                    \\ 
O       &      1.8(-4)                  & Na+     & 2.25(-9)                       \\ 
C+       &      7.86(-5)                    & P+      & 1(-9)                        \\ 
H      &      5(-5)                 & Cl+     & 2.16(-10)                        \\ 
N       &      2.47(-5)                 &   &                         \\ 
\hline
\end{tabular}
\label{tab:abundances}
$A(B) = A\times10^{B}$.
\end{table}

\section{Results} \label{sec:3}
\subsection{Comparing 1D and 3D models: D/H ratios of water}
\begin{figure*}[ht]
\resizebox{\hsize}{!}
        {\includegraphics{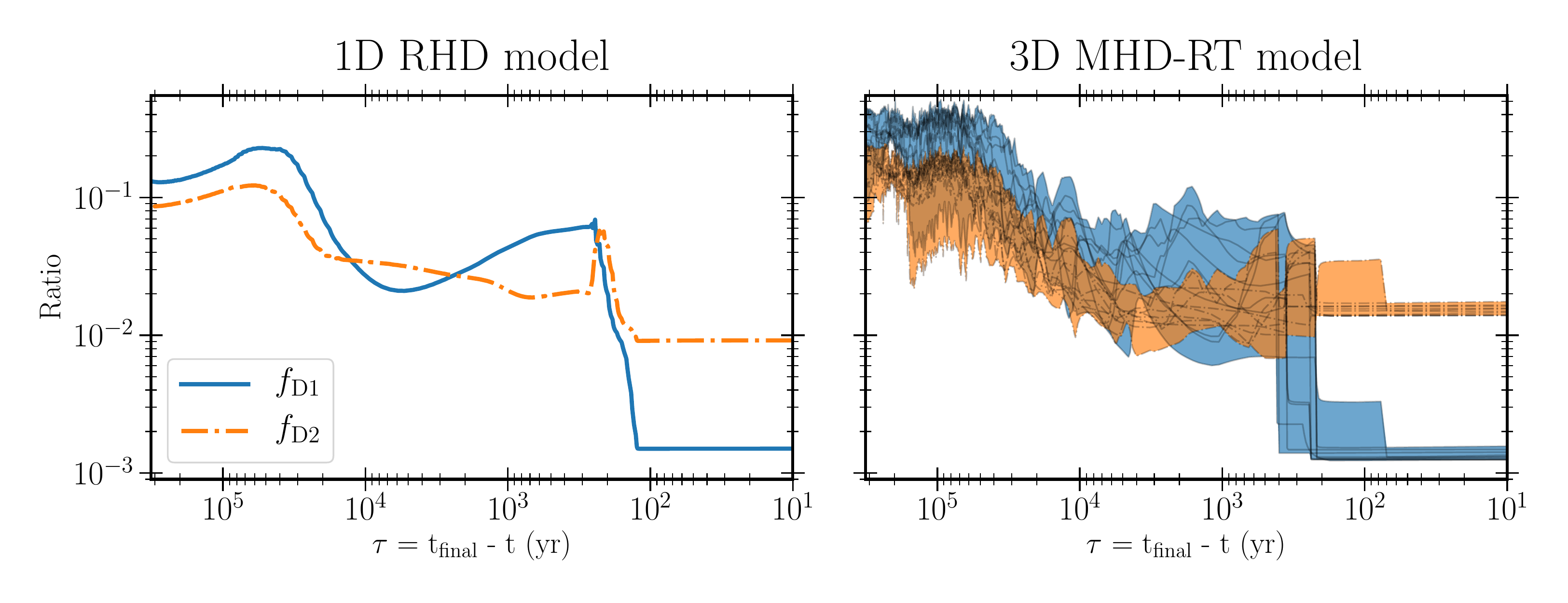}}
  \caption{Gas-phase evolution of $f_{\mathrm{D}1}$ and $f_{\mathrm{D}2}$ during protostellar collapse. The left panels shows the evolution of a particle accreted $\sim350$~kyr after core collapse toward a 1$\mathrm{M}_\odot$ protostar. The right panel shows ten particle trajectories toward the 1$\mathrm{M}_\odot$ protostars M101-C in the 3D simulations. } 
     \label{fig:1d_3d_comparison}
\end{figure*}

\begin{figure*}[ht]
\resizebox{\hsize}{!}
        {\includegraphics{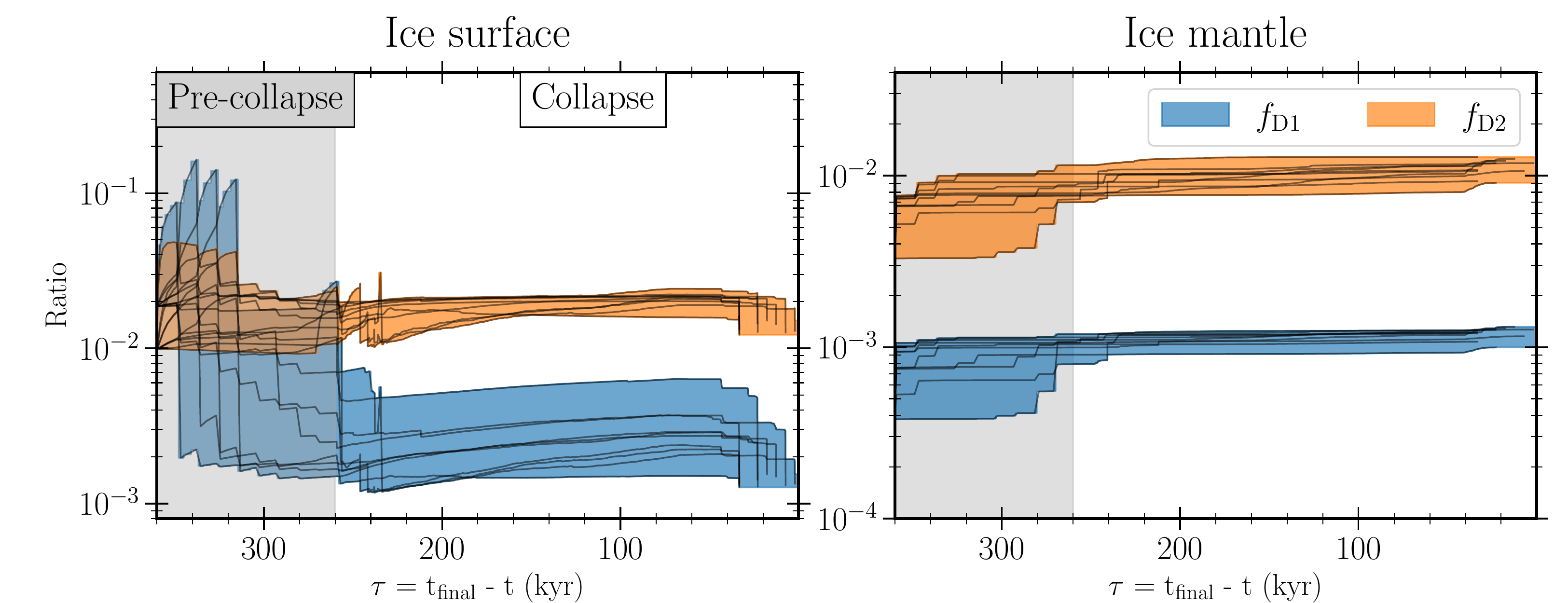}}
  \caption{Ice surface and mantle evolution for $f_{\mathrm{D}1}$ and $f_{\mathrm{D}2}$ during the protostellar collapse phase. The variation in densities for different trajectories in the pre-collapse phase induces large shifts in the ratios in the very beginning of the collapse. Once the tracer particles reach the protostellar core, the D/H ratios in the ice mantle converge to similar values, as the trajectories reach similar densities. The y-axes differ between the left and right panels.}
     \label{fig:ice_change}
\end{figure*}

To establish how the chemical evolution of water differs between 1D and 3D models of the protostellar collapse, we compare the results of the 1D RHD model presented in \citet{masunaga2000} with the 3D MHD model presented in this work. These models share the same static phase of 3~Myr as the fiducial model presented in the previous section.
In Fig. \ref{fig:1d_3d_comparison}, the evolution of gas-phase $f_{\mathrm{D}1}$ and $f_{\mathrm{D}2}$ during protostellar collapse are shown for a solar-mass protostar. The left panel shows the evolution for the 1D RHD model. The tracer particle reaches the hot corino ($T \gtrsim 100~$K) at $\tau\sim10^{2}$, where $\tau = t_\mathrm{final}-t$. Once in the hot corino, the complete desorption of the water ice lowers the D/H ratios to $f_{\mathrm{D}1}\sim10^{-3}$ and $f_{\mathrm{D}2}\sim10^{-2}$, respectively, as the surface and mantle ice D/H ratios are released and mixed with the gas-phase ratios. The right panel shows ten tracer particle trajectories from the {\sc ramses} models accreted onto sink  for the sink M101-C with a final mass of $1 \mathrm{M}_\odot$. In the figure, the shaded regions show the spread among the individual trajectories. 
In the 3D model, the gas-phase ratios show significant scatter during protostellar collapse and, generally, $f_{\mathrm{D}1}$ and $f_{\mathrm{D}2}$ are higher than in the 1D RHD model. The higher D/H ratios are  due to higher densities, which leads to more efficient gas-phase water deuteration. 
Once the particles reach the hot corino ($\tau\sim10^{3}-10^{2}$), the complete desorption of the water-ice reservoir reduces the variation drastically and the final ratios are similar for the ten tracer particles. The final D/H ratio in the gas phase is primarily set by the D/H ratio in the ice, which is similar among trajectories. 
To illustrate this, Fig. \ref{fig:ice_change} shows the change in the ice composition for the same ten particle trajectories during the pre-collapse phase (100~kyr prior to formation of the protostar) and through the collapse. The selected tracer particles all reach the hot corino at $\sim200-250$~kyr after the onset of the collapse. Initially, at $\tau \gtrsim 2.5\times10^{5}$~kyr, the D/H ratios in both surface and mantle ice show significant scatter. This scatter is caused by variation in the densities in the early evolution of the trajectories. Once the trajectories reach the protostellar core, however, the D/H ratios in the ice converge to similar values.

\begin{figure*}[ht]
\resizebox{\hsize}{!}
        {\includegraphics{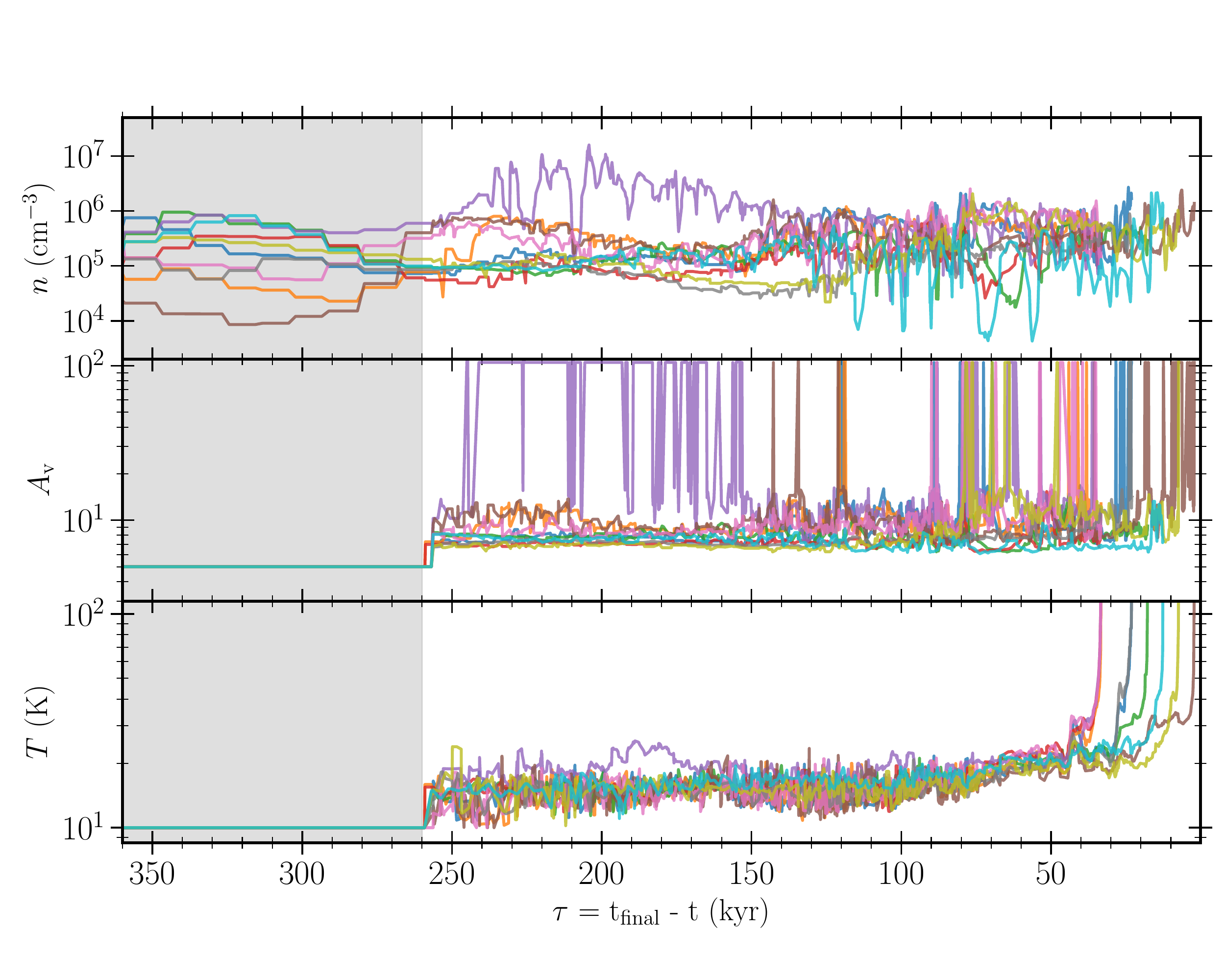}}
  \caption{Evolution in density, $A_\mathrm{v}$, and temperature for the ten tracer particles shown in Fig. \ref{fig:ice_change}. The shaded region indicates the 100~kyr pre-collapse phase, before the protostar is formed. In the pre-collapse phase, the densities are averaged in 10 kyr windows, to reduce simulation time.} 
     \label{fig:trajectories}
\end{figure*}

This change is driven by the convergence toward higher densities for the trajectories once they reach the dense core. This behavior is illustrated in Fig. \ref{fig:trajectories}, where the physical evolution of the ten particle trajectories is shown. The densities for the trajectories mainly range between 10$^{5}$--10$^6 $~$\mathrm{cm}^{-3}$ in the first 100~kyr, that is, around an order of magnitude higher than in the prestellar phase. At higher densities the thickness of the ice is increased. In the fiducial model this occurs at an epoch when water deuterium fractionation is highly efficient, which increases the D/H ratio in the water ice.
Generally, the densities within the protostellar cores are similar and this behavior is seen for all cores in the simulation, independent of mass and environment (see appendix \ref{app:radial_profiles} for the density profiles of the cores).

Toward the end of the collapse, the surface ice $f_{\mathrm{D}1}$ varies up to a factor of $\sim$4 between tracer trajectories, while $f_{\mathrm{D}2}$ varies by less than a factor of 2. Meanwhile, in the mantle reservoirs, both $f_{\mathrm{D}1}$ and $f_{\mathrm{D}2}$ are almost constant between the tracer particle trajectories. We note that while the initial variations among trajectories are lost in Fig. \ref{fig:ice_change}, variations in the physical conditions in the static phase, still impact the final D/H ratio as shown in the subsequent sections.

Overall, since the bulk ice show little variation among tracer particle trajectories in the fiducial model, a tracer particle accreted during the main accretion phase (e.g., during the Class 0 stage) shows little variation in the final D/H ratio, if we assume the same initial conditions for all tracer particles.

\subsection{Comparing isolated and clusters protostars}
\label{sec:5_2}

\begin{figure*}[ht]
\resizebox{\hsize}{!}
        {\includegraphics{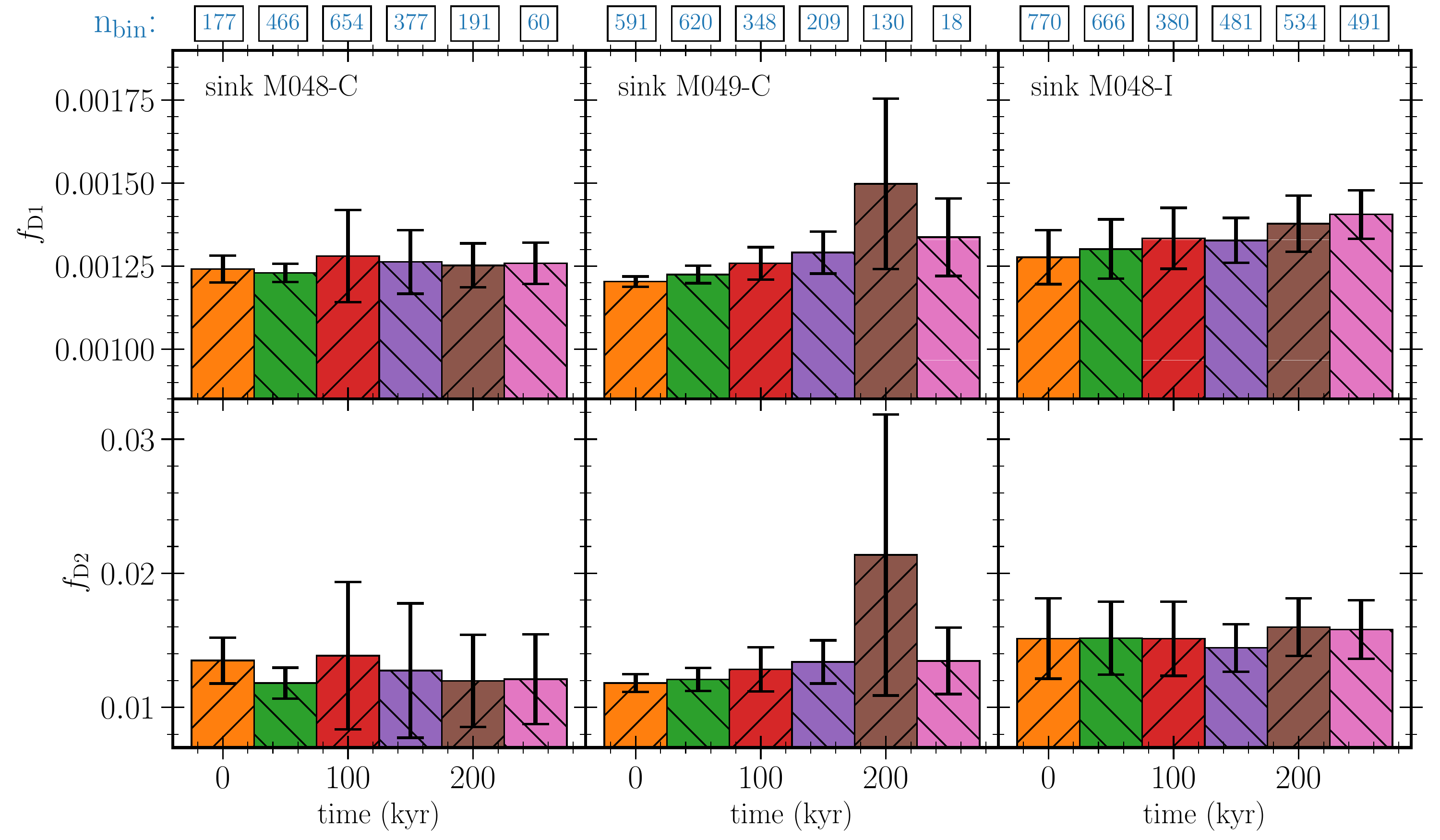}}
  \caption{HDO/H$_2$O ($f_{\mathrm{D}1}$) and D$_2$O/HDO ($f_{\mathrm{D}2}$) ratios toward three protostars in the simulation, with similar final mass of $\sim0.5 \mathrm{M}_\odot$. The tracer particles are binned according to the time at which they reach the hot corino, in bins with a width of 50 kyr. The time corresponds to the time after the onset of collapse, $t_0$. The third protostar, sink M048-I, is classified as isolated since no protostar enters within 20,000~au during the simulation, while sink M048-C and sink M049-C are clustered. Each bar shows the median values of tracer particles accreted in the time interval, and the error shows the [15.9, 84.1] percentiles. The number of tracer particles within each bin is denoted in blue above the first row.}
     \label{fig:env_comparison}
\end{figure*}

\begin{figure*}[ht]
\resizebox{\hsize}{!}
        {\includegraphics{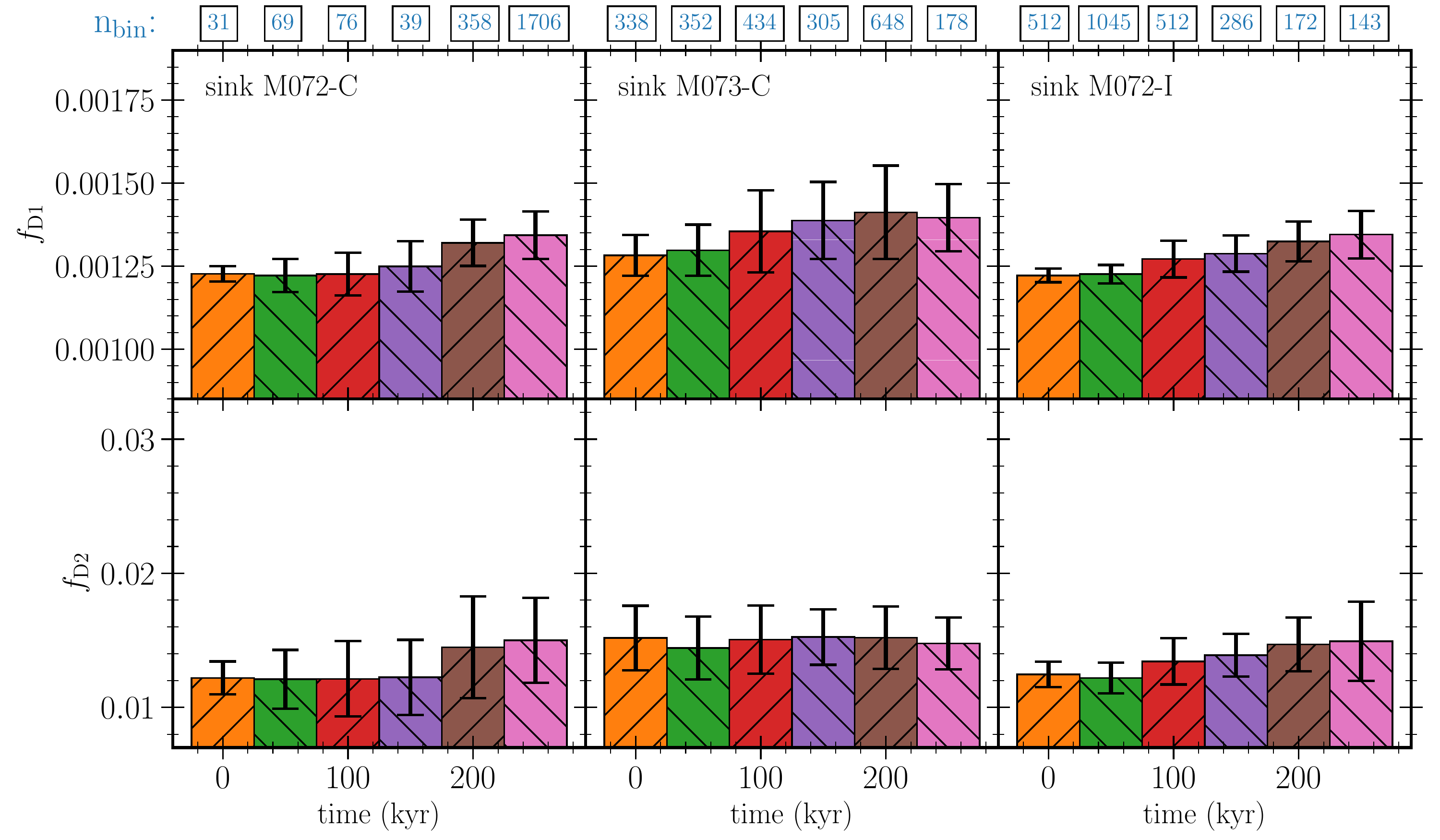}}
  \caption{Similar to Fig. \ref{fig:env_comparison}, but for protostars with final masses of $\sim0.72 \mathrm{M}_\odot$. The third protostar, sink M072-I, is classified as isolated since no protostar enters within 20,000~au during the simulation, while the remaining protostars are clustered.}
     \label{fig:env_comparison2}
\end{figure*}

In Figs. \ref{fig:env_comparison} and \ref{fig:env_comparison2}, we present comparisons between $f_{\mathrm{D}1}$ for protostars with a similar final mass. The protostars labeled M048--I and M072--I are isolated, meaning that no neighboring protostars enter within 20,000 au of the protostar for the duration of the simulation. The remaining protostars are born in clusters; there are at least two neighboring protostars within 10,000 au of the protostars at some point during the simulation. Further characteristics of the protostars are presented in Table \ref{tab:sinks}. 

The top panels show $f_{\mathrm{D}1}$ in the gas phase as the tracer particles reach the hot corino region, that is, the gas-phase HDO/H$_2$O ratio once the water ice has been sublimated off of the dust grains. We focus on this region to provide ratios directly comparable to the ratios observed toward hot corinos \citep[]{jensen2019}. 
Tracer particles are binned according to the time at which they reach the hot corino. A minor temporal evolution is evident; the tracer particles that accrete earlier showing slightly lower D/H ratios than the tracer particles accreted toward the end of the simulation, but not to a degree where the effect is observationally resolvable. As mentioned in the previous section, the bulk of the ice reservoir remains similar for all tracer particles, limiting the degree to which the D/H ratios can vary.
The bottom panels of the figures show the $f_{\mathrm{D}2}$, with an identical binning on the x-axis. $f_{\mathrm{D}2}$ shows less trend with time. This is attributed to the contemporary formation of the deuterated water isotopologs. As a longer accretion time leads to a higher overall D/H ratio, the ratio of deuterated to non-deuterated water is increased, while the ratio between the two deuterated isotopologs remain roughly constant.

No significant differences in the D/H ratios for the isolated and clustered protostars are seen in the simulations, in contrast with the observations presented in \citet{jensen2019}.

\subsection{Varying the physical conditions in the prestellar phase}
The results presented in Sect. \ref{sec:5_2} assume the same initial conditions at the onset of protostellar collapse. This assumption is unlikely to be fulfilled in dynamic star-forming regions, where temperature and radiation fields vary significantly depending on, for example, nearby massive stars and turbulence in the parental cloud. To assess the impact of the initial conditions, several combinations of initial conditions during the static phase were tested and are reported in Table \ref{tab:models}.

\subsubsection{Duration of prestellar phase}
\begin{figure*}[ht]
\resizebox{\hsize}{!}
        {\includegraphics{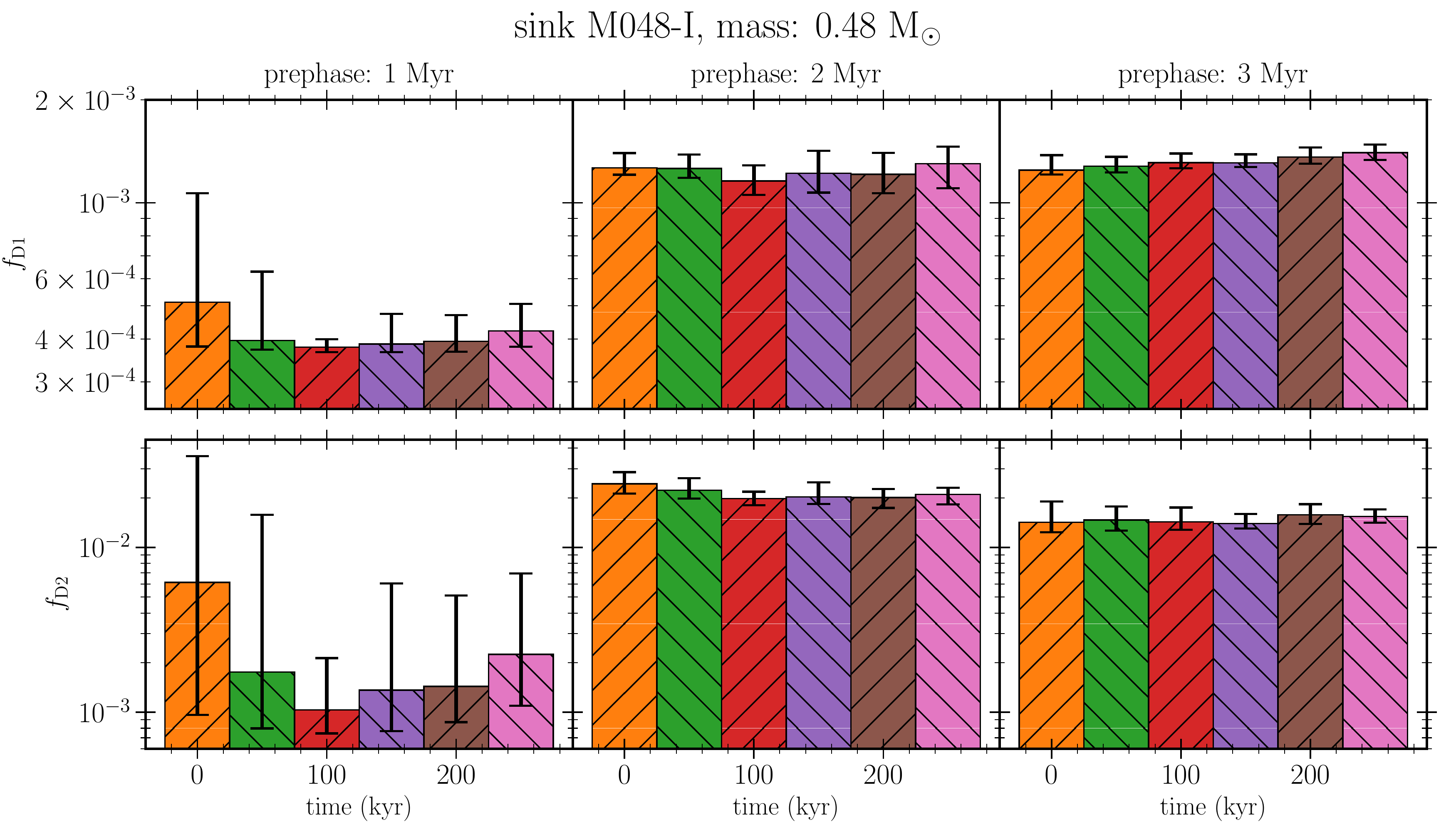}}
  \caption{$f_{\mathrm{D}1}$ and $f_{\mathrm{D}2}$ ratios toward the isolated protostar sink M048-I for varying prestellar phase durations. The tracer particles are binned according to the time at which they reach the hot corino, in bins with a width of 50 kyr. The time corresponds to the time after the onset of collapse, $t_0$. Each bar shows the median values of tracer particles accreted in the time interval, and the error shows the [15.9, 84.1] percentiles.}
     \label{fig:time_comparison126}
\end{figure*}

In Fig. \ref{fig:time_comparison126}, we show the $f_{\mathrm{D}1}$ and $f_{\mathrm{D}2}$ for static phases with durations of 1~Myr, 2~Myr, and 3~Myr. A shorter initial phase leads to a lower degree of deuterium fractionation. For a duration of 1~Myr, $f_{\mathrm{D}1} \approx 4\times10^{-4}$ and $f_{\mathrm{D}2} \approx 2\times10^{-3}$. Meanwhile, for a duration of 2~Myr, the D/H ratios are increased and the ratios are closer to the 3~Myr fiducial case, with $f_{\mathrm{D}1} \approx 1\times10^{-3}$ and $f_{\mathrm{D}2} \approx 2\times10^{-2}$. The shorter static phase of 1~Myr introduces two additional effects: an increased scatter among different particle trajectories, as indicated by the error bars in the figure, and a larger temporal variation among the tracer particles accreted early and late in the protostellar collapse. The latter effect is attributed to the shorter time to build the water ice, which has not reached a steady state at 1~Myr, making the chemical evolution more susceptible to variations in, for example, density, among the individual trajectories.

\begin{figure*}[ht]
\resizebox{\hsize}{!}
        {\includegraphics{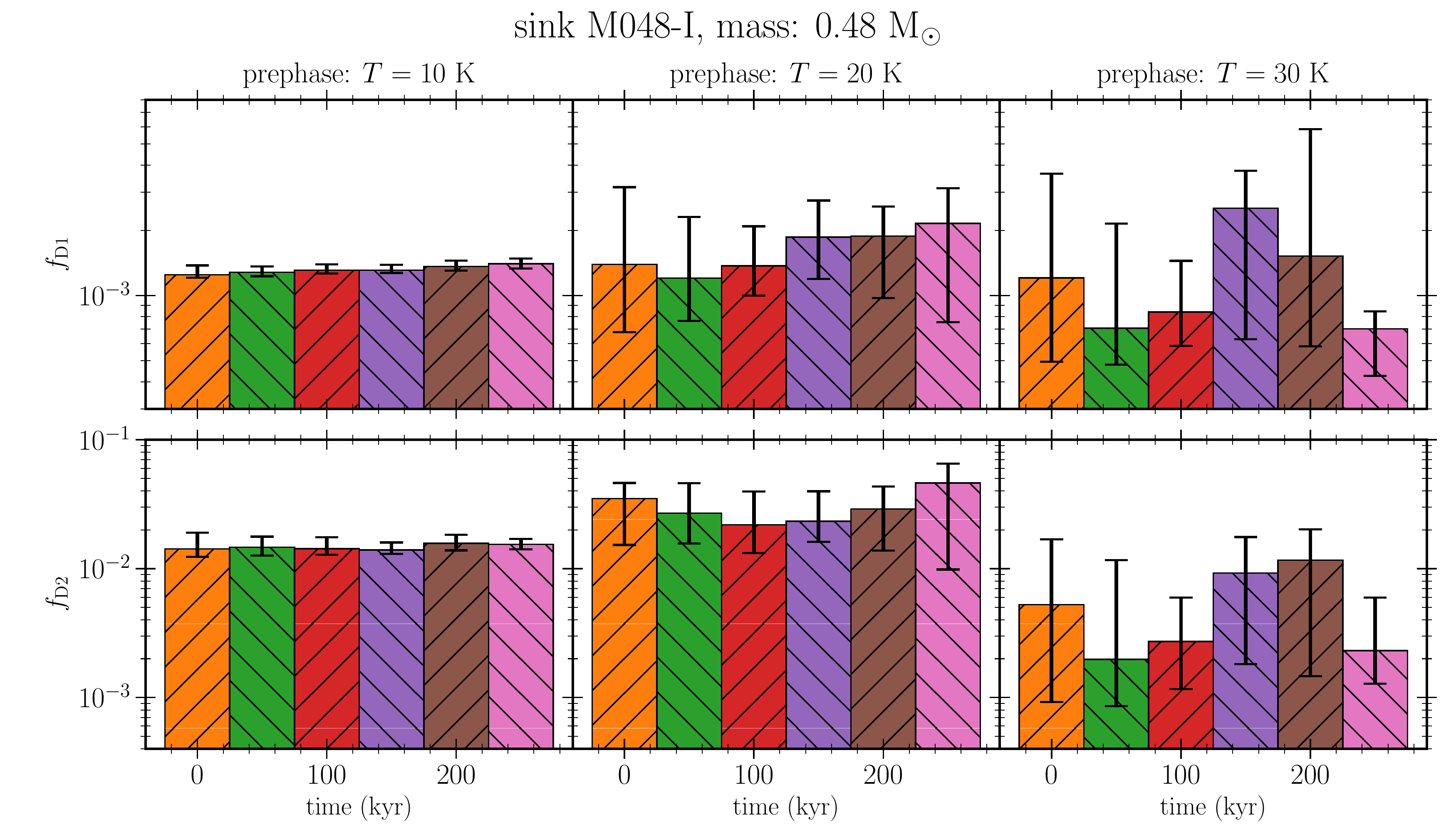}}
  \caption{$f_{\mathrm{D}1}$ and $f_{\mathrm{D}2}$ ratios toward the isolated protostar sink M048-I for varying temperatures during the prestellar phase. The tracer particles are binned according to the time at which they reach the hot corino, in bins with a width of 50 kyr. The time corresponds to the time after the onset of collapse, $t_0$. Each bar shows the median values of tracer particles accreted in the time interval, and the error shows the [15.9, 84.1] percentiles.}
     \label{fig:temp_comparison126}
\end{figure*}   

\begin{figure*}[ht]
\resizebox{\hsize}{!}
        {\includegraphics{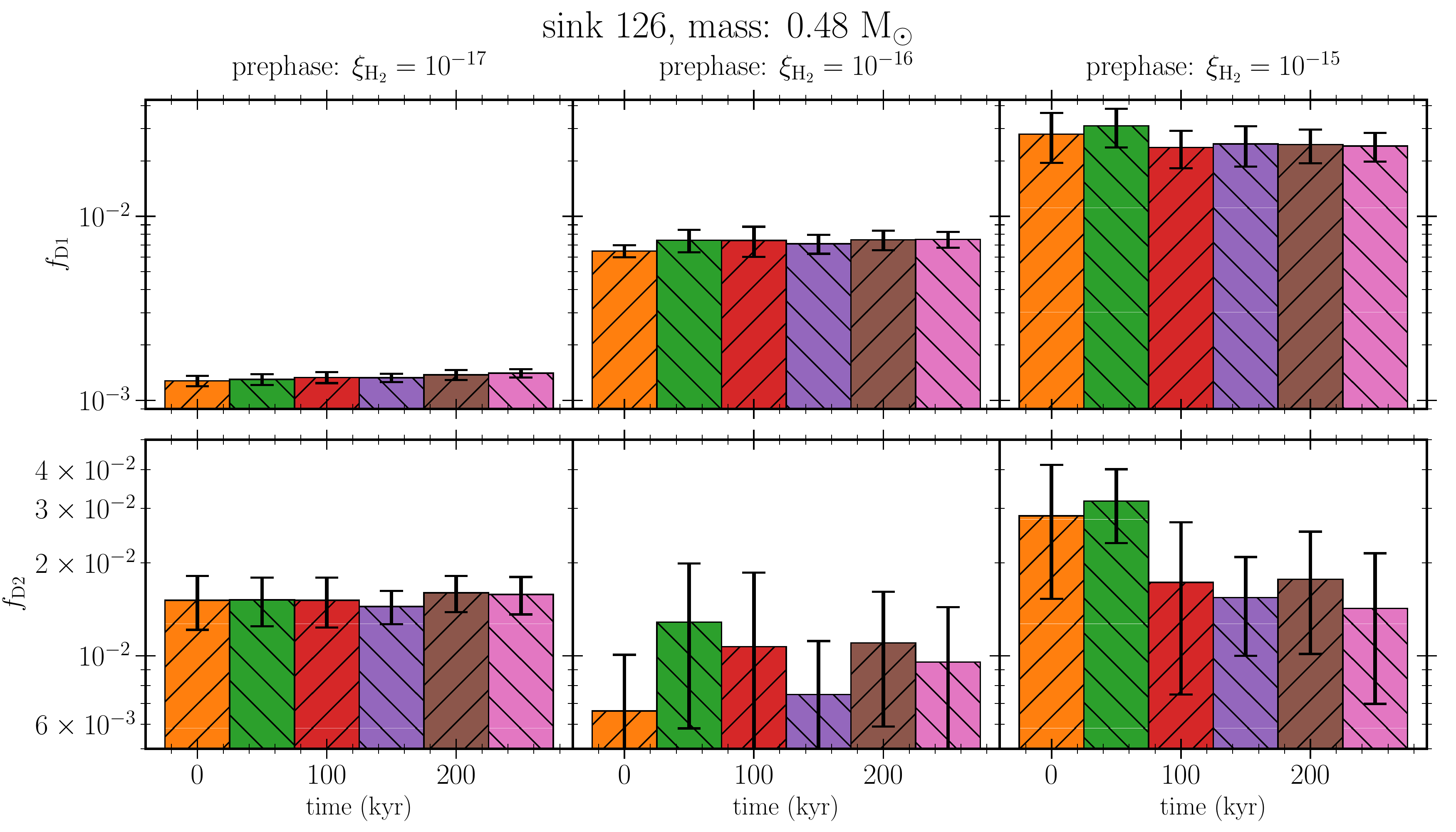}}
  \caption{$f_{\mathrm{D}1}$ and $f_{\mathrm{D}2}$ ratios toward the isolated protostar sink M048-I for cosmic-ray ionization rates $\xi_{\mathrm{H}2}$. The tracer particles are binned according to the time at which they reach the hot corino, in bins with a width of 50 kyr. The time corresponds to the time after the onset of collapse, $t_0$. }
     \label{fig:CR_comparison126}
\end{figure*}   

\begin{figure*}[ht]
\resizebox{\hsize}{!}
        {\includegraphics{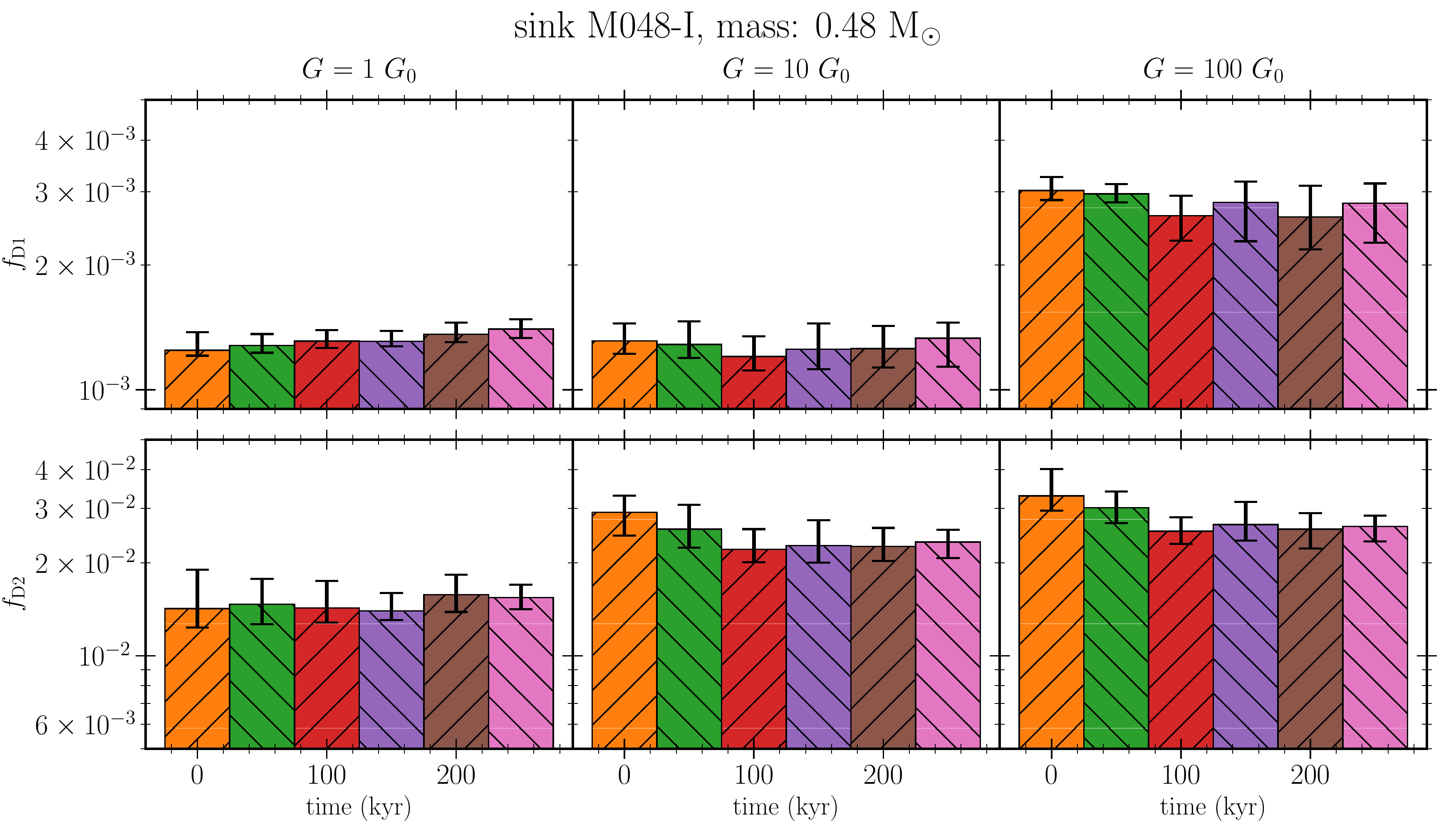}}
  \caption{$f_{\mathrm{D}1}$ and $f_{\mathrm{D}2}$ ratios toward the isolated protostar sink M048-I for varying ISRF, 1~$G_0$ corresponds to the ISRF of \citet{draine1978}. Tracer particles are binned according to the time at which they reach the hot corino, in bins with a width of 50 kyr. Time corresponds to the time after the onset of collapse, $t_0$. Each bar shows the median values of tracer particles accreted in the time interval, and the error shows the [15.9, 84.1] percentiles.}
     \label{fig:Gnot_comparison126}
\end{figure*}

\begin{table*}
\caption{Simulated D/H ratios in the hot corino for a broad range of conditions in the prestellar stage. The reported $f_{\mathrm{D}1}$ and $f_{\mathrm{D}2}$ values are the median across all tracer particles entering the hot corino.}
\centering
\begin{tabular}{llllllll}
\hline\hline
\# & $T$ (K) & Duration (Myr) & $\xi_{\mathrm{H}2}$ ($10^{-17}$s$^{-1}$) & $G_0$ & $f_{\mathrm{D}1}$ & $f_{\mathrm{D}2}$ & $\alpha$ \\\hline
1 & 10    & 1           & 1           &  1   &    $5\times10^{-4}$      &   $6\times10^{-3}$ & 12.0  \\
2 & 10    & 2           & 1           &  1   &   $1\times10^{-3}$    &   $2\times10^{-2}$ & 20.0 \\
3 & 10    & 3           & 1           &  1   &   $1\times10^{-3}$     &   $2\times10^{-2}$ & 20.0  \\
4 & 10    & 3           & 10          &  1   &    $7\times10^{-3}$      &   $1\times10^{-2}$ & 1.4 \\
5 & 10    & 3           & 100           &  1   &   $3\times10^{-2}$    &   $2\times10^{-2}$ & 0.7 \\
6 & 10    & 3           & 1           &  10   &   $1\times10^{-3}$       &   $2\times10^{-2}$  & 20.0 \\
7 & 10    & 3           & 1           &  100   &    $3\times10^{-3}$      &   $3\times10^{-2}$ & 10.0 \\
8 & 20    & 1           & 1           &  1   &    $1\times10^{-3}$      &   $1\times10^{-2}$ & 10 \\
9 & 30    & 1           & 1           &  1   &   $2\times10^{-3}$    &   $7\times10^{-3}$ & 3.5 \\
10 & 20    & 3           & 1           &  1   &    $2\times10^{-3}$      &   $3\times10^{-2}$ & 15.0 \\
11 & 30    & 3           & 1           &  1   &   $1\times10^{-3}$    &   $5\times10^{-3}$ & 5.0 \\
12 & 10    & 1.5           & 1           &  1   &   $7\times10^{-4}$    &   $8\times10^{-3}$ & 11.0 \\
13\tablefootmark{a} & 15    & 1.5           & 1           &  1   &  $7\times10^{-4}$    &   $6\times10^{-3}$ & 8.6  \\
14\tablefootmark{a} & 20    & 1.5           & 1           &  1   &   $3\times10^{-4}$    &   $2\times10^{-3}$ & 6.7 \\
\hline
\end{tabular}
\label{tab:models}
\tablefoot{
\tablefootmark{a}{For these models, the minimum temperature during the protostellar collapse was fixed at the temperature of the prestellar phase, as noted in the table.}}
\end{table*}

\subsubsection{Varying temperature}
Since the efficiency of deuterium fractionation depends on the temperature, several temperatures were tested in the prestellar phase.
In Fig. \ref{fig:temp_comparison126}, we show the results of increasing the temperature during the static phase. The radiation field and cosmic-ray flux are unaltered with respect to the fiducial model, and the prestellar duration is kept at 3~Myr. Increasing the temperature in the initial phase has only limited impact on the water D/H ratio in the hot corino for a prestellar phase. While a higher temperature in the prestellar phase reduces the $f_{\mathrm{D}1}$ and $f_{\mathrm{D}2}$ in the ice prior to collapse, the 100~kyr pre-collapse phase where the temperature is fixed at 10~K followed by low temperatures during the early phases of the protostellar collapse are sufficient to increase the $f_{\mathrm{D}1}$ and $f_{\mathrm{D}2}$ to $\gtrsim10^{-3}$. The efficient deuterium fractionation on these short timescales is a consequence of the high densities in the cores ($n_\mathrm{H} \gtrsim 10^{5}~\mathrm{cm}^{-3}$). For models $\#13$ and $\#14$ listed in Table \ref{tab:models}, minimum temperatures of 15~K and 20~K, respectively, were enforced during the protostellar collapse. In these cases, the final D/H ratios in the hot corino remain lower, because the deuterium fractionation processes are less efficient throughout the entire chemical evolution. A stronger impact of higher temperatures is expected at temperatures > 30~K, or at lower densities, where CO freeze-out is less efficient. 

\subsubsection{Varying ISRF and cosmis-ray ionization rate}
Increasing the cosmic-ray ionization rate leads to a higher $f_{\mathrm{D}1}$, from $\sim10^{-3}$ at $\xi_{\mathrm{H}2} = 10^{-17}~\mathrm{s}^{-1}$, to $f_{\mathrm{D}1} > 10^{-2}$ at $\xi_{\mathrm{H}2} = 10^{-15} ~\mathrm{s}^{-1}$. The impact on $f_{\mathrm{D}2}$ is more complex. At $\xi_{\mathrm{H}2} = 10^{-16}~\mathrm{s}^{-1}$, $f_{\mathrm{D}2}$ is slightly lower than the fiducial case, while at $\xi_{\mathrm{H}2} = 10^{-15}~\mathrm{s}^{-1}$, $f_{\mathrm{D}2}$ is higher than the fiducial model early in the collapse and similar toward the end. Higher cosmic-ray ionization rates lead to a larger spread in the D$_2$O/HDO ratios between particles trajectories, as seen in Fig. \ref{fig:CR_comparison126}. The increased deuterium fractionation at stronger cosmic-ray fluxes is driven by the formation of $\mathrm{H}_{3}^{+}$ and deuterated isotopologs, which are central to the fractionation processes in the cold ISM. The formation rates of $\mathrm{H}_{3}^{+}$ and isotopologs are regulated by the cosmic-ray ionization of H$_2$ and consequently an increased cosmic-ray flux can increase rate of the fractionation process. Furthermore, higher ionization rates reduce the ortho-to-para conversion timescale for H$_2$, which further enhances deuteration. The different formation pathways for HDO and D$_2$O (i.e., HDO formation through surface reactions and D$_2$O formation in the gas-phase) leads to the discrepancy between $f_{\mathrm{D}1}$ and $f_{\mathrm{D}2}$, as the efficiencies of the formation pathways differ.

Increasing the ISRF to 10~$G_0$ in the static phase leads to a slight increase of $f_{\mathrm{D}2}$, while $f_{\mathrm{D}1}$ remains roughly constant. At $G = 100~G_0$, $f_{\mathrm{D}1}$ and $f_{\mathrm{D}2}$ are increased compared with the fiducial model. These effects are due to different branching ratios for photodissociation of HDO, where the branch producing H + OD is favored over the D + OH branch. This leads to more efficient reformation of deuterated water, increasing deuterium fractionation through photodissociation in both gas-phase and grain-surface chemistry \citep[see, e.g., ][ and references therein]{arasa2015}. 
We note that a visual extinction of $A_\mathrm{v} = 5$~mag was kept during the testing of initial conditions, and the ISRF hence remained heavily attenuated. At a lower visual extinction, stronger radiation fields lower the D/H ratio, as CO freeze-out is reduced by photodesorption.


\subsubsection{Reproducing observed HDO/H$_2$O and D$_2$O/HDO ratios in hot corinos}
Current estimates of $f_{\mathrm{D}1}$ and $f_{\mathrm{D}2}$ in hot corinos of low-mass protostars are limited to a few sources. \citet{coutens2014} detect D$_2$O toward IRAS~2A and derive $\alpha = f_{\mathrm{D}2}/f_{\mathrm{D}1} \gtrsim 7$. Similar high values of $\alpha$ are also seen for other protostars (S. Jensen et al., subm.). 
We require $\alpha \gtrsim 5$ and $f_{\mathrm{D}1}$ in the range $10^{-4}-10^{-3}$ to reproduce current observations of the water D/H ratios in hot corinos \citep{persson2014, jensen2019}.
The fiducial model reproduced the observed $f_{\mathrm{D}1} \sim 10^{-3}$ toward isolated protostars presented in \citet{jensen2019}, while the lower D/H ratio toward the clustered protostars require different initial conditions. With the model presented in this work, reproducing $10^{-4} < f_{\mathrm{D}1} < 10^{-3}$ requires a shorter prestellar core phase or higher temperatures during the collapse. Model \#12, featuring a 1.5~Myr prestellar phase, yields $f_{\mathrm{D}1} \sim 7\times10^{-4}$ and $\alpha \approx 13$, within the observed range for clustered sources. Similarly, models \#13 and \#14, where minimum temperatures of 15~K and 20~K are enforced during the entire collapse, yield $f_{\mathrm{D}1} \sim 7\times10^{-4}$ and $f_{\mathrm{D}1} \sim 3\times10^{-4}$, respectively, which is well within the observed ranges.  
These results are in line with the predictions from \citet{jensen2019}, who proposed that shorter timescale or higher temperatures could drive the observed differentiation between isolated and clustered protostars.
Increasing the cosmic-ray flux changes $\alpha$ substantially, as shown in Table \ref{tab:models}, and the observed D/H ratios toward low-mass Class 0 protostars appear to be incompatible with higher cosmic-ray fluxes ($\xi_{\mathrm{H}_2} = 10^{-15}$). The impact of a stronger ISRF on the D/H ratios of water is limited within the model, where the visual extinction is relatively high during the entire simulation ($A_\mathrm{v} \gtrsim 5$). 



\section{Discussion}  \label{sec:4}
\subsection{Impact of the environment during protostellar collapse}
The comparison of water D/H ratios in realistic 3D protostellar collapse models show no significant variation between clustered and isolated protostars. The mass of the protostar does not appear to influence the water chemistry notably either. The homogenity of the D/H ratios results from several factors. First, the physical structure of the low-mass protostellar cores presented in this work are similar; the density profiles resemble Bonnor-Ebert spheres at the onset of collapse \footnote{Radial density profiles for the protostars are presented in Appendix \ref{app:radial_profiles}}, and once a tracer particle reaches the core, the D/H ratios in the ice converge, independent of the surrounding environment. Second, the identical initial conditions in the model prior to the protostellar collapse limit the possible range of D/H in the hot corino. Our fiducial model includes a 3~Myr static core phase, during which the water D/H ratio is established in the ice. Throughout the protostellar collapse, variations in tracer trajectories (e.g., $n(t)$ and $T(t)$) can induce variations in the gas-phase D/H ratios, but the bulk water reservoir shows limited variation between trajectories and once the thick ice mantle is evaporated, the gas-phase D/H ratio is largely similar to the ice D/H ratio which is set by the physical conditions during ice formation (i.e., the evolution in the molecular cloud and static core). 

While the observed differentiation in water deuteration toward isolated and clustered low-mass protostars is not explained by chemical processing during the protostellar collapse, the observed feature can be explained by variations in the physical evolution prior to protostellar collapse. A shorter static phase of $\lesssim1.5$~Myr results in $f_{\mathrm{D}1} \lesssim 7\times10^{-4}$, which agrees with the results toward clustered sources \citep{persson2014}. Furthermore, these conditions produce $f_{\mathrm{D}2} \sim10^{-2}$, yielding $\alpha \gtrsim 5$ as observed toward IRAS~2A \citep{coutens2014}. Increasing the temperature to 20~K during the entire collapse leads to a lower $f_{\mathrm{D}1} \sim 3\times10^{-4}$, while $\alpha \sim 7$ is reproduced for a prephase duration of $> 1.5$~Myr. Overall, temperatures in the range $T \lesssim 20$~K can reproduce the observed D/H ratios.

These models do not self-consistently predict the physical and chemical evolution during the molecular cloud and static core phase, however, both a shorter static phase and higher temperatures are feasible variations in dynamic molecular clouds. A shorter collapse timescale for clustered sources can be driven by higher densities or through external forces acceleration the collapse \citep[e.g.,][]{ward-thompson2007, enoch2008}. Similarly, higher temperatures in clustered regions can occur through irradiation from nearby massive protostars or through shock heating in turbulent cloud environments \citep[e.g.,][]{krumholz2014}.

\subsection{Caveats of the physical models}\label{sec:4_2}
Variations in cloud conditions prior to the protostellar collapse are not included in the models presented, except for a 100~kyr period prior to the protostellar collapse where the temperature is fixed at 10~K. Variations in the initial conditions are necessary to reproduce the observed variation in D/H ratios among clustered and isolated protostars. To self-consistently model the conditions in the molecular cloud and static core phase, several enhancements to the model setup are required. Currently, radiative transfer is calculated locally around protostars once the protostellar collapse is initiated. The temperature in the gas prior to collapse is therefore unknown. Implementing global radiative transfer coupled with a thermochemical code, could provide better constraints on the temperature and radiation field during the molecular cloud and static phases \citep[e.g.,][]{glover2010, grassi2014}.
To assess whether the fixed temperature of 10~K during the 100~kyr pre-collapse phase influenced the results, we reran the fiducial models with variable temperatures and extinctions derived from parametric relations during this phase.
The extinction was derived from the total number density through the relation:
\begin{equation}
    A_\mathrm{v} = \Big(\frac{n_{\mathrm{H}}}{10^3 \mathrm{cm}^{-3} }\Big)^{2/3} ~\mathrm{mag} \, .
\end{equation}
The above relation is derived in \citet{grassi2017} by fitting a power law to the density-extinction relation in the RHD models of \citet{glover2010}.
A semi-analytical relation between extinction and temperature is derived and tested in \citet{hocuk2017} as follows:
\begin{equation}
    T_\mathrm{dust} = 11 + 5.7\times\tanh[0.61 - \log_{10}(A_\mathrm{v})] \, ~\mathrm{K} .
\end{equation}
Implementing these relations during the pre-collapse phase had minimal effect on the final D/H ratios and did not influence the conclusions presented in this work. This emphasizes that the D/H ratio is predominantly a product of the initial conditions, with limited variation among different tracer particle trajectories.


The {\sc ramses} models are limited to by the grid resolution of 25 au. The resolution limit means that neither circumstellar disks nor outflows are sufficiently resolved. Resolving the physical structure of both outflow cavities and the disk is necessary to accurately predict the radiation and temperature field in the later stages of the protostellar formation. We are therefore limited to studying the initial collapse phase, where most of the matter is accreted. Increasing the resolution allows the disk and outflow cavities to be resolved and the results presented in \citet{kuffmeier2018} demonstrate that disks are formed early on around the Class 0/I stage.

The three-phase model presented in this work does not consider bulk diffusion, that is, diffusion between surface and mantle or reactions within the ice mantle. Including bulk diffusion may impact the chemical evolution of three-phase models, especially during a protostellar collapse \citep{garrod2013}. Without bulk diffusion, molecules can remain locked in the ice mantle at temperatures exceeding their sublimation temperature, since only the surface molecules are free to sublimate. For instance, CO remains in the ice mantle at temperatures above $\sim$30~K, since only the surface CO can desorb to the gas phase. This can alter the evolution of COMs and the change in CO gas-phase abundance; however, this does not impact the gas-phase deuterium fractionation process notable \citep{furuya2015}. 
\citet{furuya2017} study the evolution of HDO/H$_2$O and D$_2$O/HDO during protostellar collapse and find that the inclusion of bulk-diffusion does not impact the D/H ratio of water post-collapse. Owing to the high binding energy of water, bulk diffusion of water is likely inefficient and the effect is unlikely to alter the strong inheritance shown in the models.

\subsection{Inheritance of water D/H ratio}
Determining whether the D/H ratio is inherited or reset during star and planet formation is critical for our understanding of the evolution and delivery of water to young planets \citep[e.g.,][]{dishoeck2014}.
Through simulations of water deuterium fractionation in a protoplanetary disk, \citet{cleeves2014} showed that cometary D/H ratios are too high to originate from local processing within the disk. Hence, some degree of inheritance is necessary to reproduce the deuterium fractionation in the Solar System, including the Earth's oceans. In a later work, \citet{furuya2017} study the evolution of the D/H ratio of water, combining 1D protostellar collapse models with a 2D axisymmetric protoplanetary disk model. These authors find that the majority of the water in the protoplanetary disk remained unprocessed, again favoring inheritance for the majority of the water present in the disk. 
The results presented in this work favor chemical inheritance for the D/H ratio of water in the hot corino as opposed to a ``reset'' chemistry post-collapse. Variation in physical evolution during the protostellar collapse has little impact on the degree of water deuteration in the hot corino, which is dominated by the conditions in the static phase. However, the D/H ratio set in the static phase is not fixed and may evolve as the gas and dust enters the protostellar core, where densities increase. While this process adjusts the D/H ratio in the ice, it is not a source of differentiation in the final D/H ratios for water in the hot corinos since the structures of low-mass protostellar cores are largely similar between isolated and clustered protostars, owing to the physical processes that dominate the stability of cloud cores \citep[e.g.,][]{kuffmeier2017}. 

\citet{furuya2017} suggest that $\alpha$ is a better tool for distinguishing between inheritance and resetting of the deuterium fractionation, since their simulations yield $\alpha > 1$ under prestellar conditions, while resetting of the chemistry in the protoplanetary disk leads to $\alpha < 1$. Our results confirm that $\alpha > 1$ in the case of inherited D/H from the static phase when standard ISRF and cosmic-ray ionization rates are considered. This further motivates the study of D$_2$O/HDO ratios in hot corino to provide further constrains on the chemical evolution of water from cloud to core.

If the deuterium fractionation of water in the hot corino is inherited, the D/H ratio of water at later stages in the star and planet formation process could be linked to the initial conditions in the cloud environment prior to the onset of star formation. However, the nature of hot corinos, and the chemical link from hot corino to protoplanetary disk is not well established at this stage \citep[e.g., ][]{belloche2020, jorgensen2020}. In a recent work, \citet{coutens2020} study the chemical evolution from the collapse of a prestellar core to the formation of a protoplanetary disk. As in this work, the physical evolution during the collapse is tracked using tracer particles. Their model couples a 3D~non-ideal MHD {\sc ramses} simulation of a collapsing core with the {\sc nautilus} chemical model and focuses on a broad range of molecules. These authors find a clear chemical link between the chemical composition in the prestellar core and in the protoplanetary disk. Generally, the principal elemental carriers remain the same throughout the collapse, with the exception of sulfur and phosphorus, where the main reservoirs evolve. Furthermore, the abundances of the most abundant molecules (e.g., H$_2$O, CH$_4$, and CH$_3$OH) remain largely unaltered through the collapse, although the abundance of ``large'' COMs (e.g., CH$_3$CHO, CH$_3$OCH$_3$) increases during the warm-up phase. Overall, these conclusions agree with the results presented in this work for water, namely, that the initial conditions have a high impact on the final chemical composition; this supports a high degree of inheritance during the protostellar collapse. 

In recent years, rotational signatures have been observed in a limited sample of young embedded Class 0 sources \citep[e.g., ][]{tobin2012, murillo2013}, including one source hosting a hot corino \citep[]{bjerkeli2019, imai2019}. This could suggest that the chemistry observed in hot corinos reflects the same chemical reservoir that constitutes the initial disk structure. Still, detections of the warm gas component in sources with clear disk structures remains elusive \citep[e.g.,][]{villarmois2018}. Furthermore, once the disk is formed, in situ chemical processes may locally alter the chemical composition throughout the disk, introducing chemical gradients in the radial and vertical directions \citep[e.g., ][]{henning2013}. 
Recently, a correlation between COMs in the hot corino of IRAS16293--2422 and the comet 67P were established, supporting a potential link between the chemical composition of the hot corino and protoplanetary disk, where comets are formed \citep{maria2019}. This could imply limited chemical processing in the protosolar nebula at larger orbital distances where comets were formed. Furthermore, a recent study of COMs in the disk around V883 Ori found similarities with the chemistry in hot corinos and comets \citep{lee2019}. 

\section{Summary}  \label{sec:5}
In this work we present physicochemical models of protostellar collapse to study how variations in protostellar cloud environments may impact the chemistry, specifically the D/H ratio of water. 
The main results are as follows:
\begin{itemize}
    \item The HDO/H$_2$O and D$_2$O/HDO ratios show no clear correlation with the cloud environment. Physical variations during the protostellar collapse are not sufficient to drive a notable difference in the deuterium fractionation between different protostars as the D/H ratio is governed by the initial cloud conditions prior to protostellar collapse. 
    \item To reproduce the observed differentiation in HDO/H$_2$O ratios between isolated and clustered protostar, variations in the duration or temperature of the prestellar phase are necessary. As such, the D/H ratio is determined by these conditions, favoring inheritance of the water D/H ratio observed in hot corinos. Static phases with durations of 1--3~Myr and temperatures 10--20~K can reproduce the observed HDO/H$_2$O and D$_2$O/HDO ratios.
    \item Increasing the cosmic-ray ionization rates to $\xi_{\mathrm{H}2} \approx 10^{-15}~\mathrm{s}^{-1}$ leads to D/H ratios that are incompatible with the observed values in hot corinos toward low-mass Class 0 sources.
    \item Global chemical models of star-forming regions are needed to establish the origin of chemical differentiation in star-forming regions. Tracing the chemical evolution from molecular cloud to hot corino is a necessary step to self-consistently determine which processes drive chemical differentiation of water and more complex molecules.
\end{itemize}

This work illustrates how isotope fractionation can trace the chemical evolution through the star and planet formation process, by combining detailed models with the latest observational advances. Furthermore, present results highlight the need for physicochemical models spanning the entire star formation process, to accurately model the origin and evolution of chemical complexity during the formation of stars and planets. 

\begin{acknowledgements}
    The authors are grateful to Tommaso Grassi, who provided the underlying framework for the chemical model. The authors would like to thank the referee for constructive comments which helped improve the manuscript.
	The group of JKJ acknowledges support from the European Research Council (ERC) under the European Union's Horizon 2020 research and innovation programme (grant agreement No 646908) through ERC Consolidator Grant "S4F". TH was supported by the Independent Research Fund Denmark through grant No. DFF 8021--00350B. We acknowledge PRACE for awarding us access to Curie at GENCI@CEA, France, and to Marconi at CINECA, Italy used to carry out the simulation. Resources at the University of Copenhagen HPC centre, funded in part by the Carlsberg, Novo, and Villum foundations, were used for the data analysis. This work made use of {\sc matplotlib} \citep{hunter2007} and {\sc numpy} \citep{2020NumPy-Array}.
\end{acknowledgements}

%
%

\bibliographystyle{aa}
\bibliography{water_models.bib}

\begin{thebibliography}{89}
\expandafter\ifx\csname natexlab\endcsname\relax\def\natexlab#1{#1}\fi

\bibitem[{{Aikawa} {et~al.}(2020){Aikawa}, {Furuya}, {Yamamoto}, \&
  {Sakai}}]{aikawa2020}
{Aikawa}, Y., {Furuya}, K., {Yamamoto}, S., \& {Sakai}, N. 2020, \apj, 897, 110

\bibitem[{{Aikawa} {et~al.}(2001){Aikawa}, {Ohashi}, {Inutsuka}, {Herbst}, \&
  {Takakuwa}}]{aikawa2001}
{Aikawa}, Y., {Ohashi}, N., {Inutsuka}, S.-i., {Herbst}, E., \& {Takakuwa}, S.
  2001, \apj, 552, 639

\bibitem[{{Aikawa} {et~al.}(2012){Aikawa}, {Wakelam}, {Hersant}, {Garrod}, \&
  {Herbst}}]{aikawa2012}
{Aikawa}, Y., {Wakelam}, V., {Hersant}, F., {Garrod}, R.~T., \& {Herbst}, E.
  2012, \apj, 760, 40

\bibitem[{{Altwegg} {et~al.}(2017){Altwegg}, {Balsiger}, {Berthelier},
  {Bieler}, {Calmonte}, {De Keyser}, {Fiethe}, {Fuselier}, {Gasc}, {Gombosi},
  {Owen}, {Le Roy}, {Rubin}, {S{\'e}mon}, \& {Tzou}}]{altwegg2017}
{Altwegg}, K., {Balsiger}, H., {Berthelier}, J.~J., {et~al.} 2017,
  Philosophical Transactions of the Royal Society of London Series A, 375,
  20160253

\bibitem[{{Arasa} {et~al.}(2015){Arasa}, {Koning}, {Kroes}, {Walsh}, \& {van
  Dishoeck}}]{arasa2015}
{Arasa}, C., {Koning}, J., {Kroes}, G.-J., {Walsh}, C., \& {van Dishoeck},
  E.~F. 2015, \aap, 575, A121

\bibitem[{{Artur de la Villarmois} {et~al.}(2018){Artur de la Villarmois},
  {Kristensen}, {J{\o}rgensen}, {Bergin}, {Brinch}, {Frimann}, {Harsono},
  {Sakai}, \& {Yamamoto}}]{villarmois2018}
{Artur de la Villarmois}, E., {Kristensen}, L.~E., {J{\o}rgensen}, J.~K.,
  {et~al.} 2018, \aap, 614, A26

\bibitem[{{Belloche} {et~al.}(2020){Belloche}, {Maury}, {Maret}, {Anderl},
  {Bacmann}, {Andr{\'e}}, {Bontemps}, {Cabrit}, {Codella}, {Gaudel}, {Gueth},
  {Lef{\`e}vre}, {Lefloch}, {Podio}, \& {Testi}}]{belloche2020}
{Belloche}, A., {Maury}, A.~J., {Maret}, S., {et~al.} 2020, \aap, 635, A198

\bibitem[{{Bjerkeli} {et~al.}(2019){Bjerkeli}, {Ramsey}, {Harsono}, {Calcutt},
  {Kristensen}, {van der Wiel}, {J{\o}rgensen}, {Muller}, \&
  {Persson}}]{bjerkeli2019}
{Bjerkeli}, P., {Ramsey}, J.~P., {Harsono}, D., {et~al.} 2019, \aap, 631, A64

\bibitem[{{Bohlin} {et~al.}(1978){Bohlin}, {Savage}, \& {Drake}}]{bohlin1978}
{Bohlin}, R.~C., {Savage}, B.~D., \& {Drake}, J.~F. 1978, \apj, 224, 132

\bibitem[{{Caselli} {et~al.}(2012){Caselli}, {Keto}, {Bergin}, {Tafalla},
  {Aikawa}, {Douglas}, {Pagani}, {Y{\'\i}ld{\'\i}z}, {van der Tak}, {Walmsley},
  {Codella}, {Nisini}, {Kristensen}, \& {van Dishoeck}}]{caselli2012}
{Caselli}, P., {Keto}, E., {Bergin}, E.~A., {et~al.} 2012, \apj, 759, L37

\bibitem[{{Caselli} {et~al.}(2002){Caselli}, {Stantcheva}, {Shalabiea},
  {Shematovich}, \& {Herbst}}]{caselli2002}
{Caselli}, P., {Stantcheva}, T., {Shalabiea}, O., {Shematovich}, V.~I., \&
  {Herbst}, E. 2002, \planss, 50, 1257

\bibitem[{{Caselli} {et~al.}(2003){Caselli}, {van der Tak}, {Ceccarelli}, \&
  {Bacmann}}]{caselli2003}
{Caselli}, P., {van der Tak}, F.~F.~S., {Ceccarelli}, C., \& {Bacmann}, A.
  2003, \aap, 403, L37

\bibitem[{{Chuang} {et~al.}(2016){Chuang}, {Fedoseev}, {Ioppolo}, {van
  Dishoeck}, \& {Linnartz}}]{chuang2016}
{Chuang}, K.~J., {Fedoseev}, G., {Ioppolo}, S., {van Dishoeck}, E.~F., \&
  {Linnartz}, H. 2016, \mnras, 455, 1702

\bibitem[{{Cleeves} {et~al.}(2014){Cleeves}, {Bergin}, {Alexand er}, {Du},
  {Graninger}, {{\"O}berg}, \& {Harries}}]{cleeves2014}
{Cleeves}, L.~I., {Bergin}, E.~A., {Alexand er}, C. M.~O.~D., {et~al.} 2014,
  Science, 345, 1590

\bibitem[{{Coutens} {et~al.}(2020){Coutens}, {Commer{\c{c}}on}, \&
  {Wakelam}}]{coutens2020}
{Coutens}, A., {Commer{\c{c}}on}, B., \& {Wakelam}, V. 2020, \aap, 643, A108

\bibitem[{{Coutens} {et~al.}(2014){Coutens}, {J{\o}rgensen}, {Persson}, {van
  Dishoeck}, {Vastel}, \& {Taquet}}]{coutens2014}
{Coutens}, A., {J{\o}rgensen}, J.~K., {Persson}, M.~V., {et~al.} 2014, \apj,
  792, L5

\bibitem[{{Coutens} {et~al.}(2013){Coutens}, {Vastel}, {Cabrit}, {Codella},
  {Kristensen}, {Ceccarelli}, {van Dishoeck}, {Boogert}, {Bottinelli},
  {Castets}, {Caux}, {Comito}, {Demyk}, {Herpin}, {Lefloch}, {McCoey},
  {Mottram}, {Parise}, {Taquet}, {van der Tak}, {Visser}, \&
  {Y{\i}ld{\i}z}}]{coutens2013}
{Coutens}, A., {Vastel}, C., {Cabrit}, S., {et~al.} 2013, \aap, 560, A39

\bibitem[{{Cuppen} {et~al.}(2017){Cuppen}, {Walsh}, {Lamberts}, {Semenov},
  {Garrod}, {Penteado}, \& {Ioppolo}}]{cuppen2017}
{Cuppen}, H.~M., {Walsh}, C., {Lamberts}, T., {et~al.} 2017, \ssr, 212, 1

\bibitem[{{D'Antona} \& {Mazzitelli}(1997)}]{pms1997}
{D'Antona}, F. \& {Mazzitelli}, I. 1997, \memsai, 68, 807

\bibitem[{{Diplas} \& {Savage}(1994)}]{diplas1994}
{Diplas}, A. \& {Savage}, B.~D. 1994, \apj, 427, 274

\bibitem[{{Draine}(1978)}]{draine1978}
{Draine}, B.~T. 1978, \apjs, 36, 595

\bibitem[{{Draine} \& {Bertoldi}(1996)}]{draine1996}
{Draine}, B.~T. \& {Bertoldi}, F. 1996, \apj, 468, 269

\bibitem[{{Drozdovskaya} {et~al.}(2019){Drozdovskaya}, {van Dishoeck}, {Rubin},
  {J{\o}rgensen}, \& {Altwegg}}]{maria2019}
{Drozdovskaya}, M.~N., {van Dishoeck}, E.~F., {Rubin}, M., {J{\o}rgensen},
  J.~K., \& {Altwegg}, K. 2019, \mnras, 490, 50

\bibitem[{{Dullemond} {et~al.}(2012){Dullemond}, {Juhasz}, {Pohl}, {Sereshti},
  {Shetty}, {Peters}, {Commercon}, \& {Flock}}]{dullemond2012}
{Dullemond}, C.~P., {Juhasz}, A., {Pohl}, A., {et~al.} 2012, {RADMC-3D: A
  multi-purpose radiative transfer tool}

\bibitem[{{Enoch} {et~al.}(2008){Enoch}, {Evans}, {Sargent}, {Glenn},
  {Rosolowsky}, \& {Myers}}]{enoch2008}
{Enoch}, M.~L., {Evans}, Neal~J., I., {Sargent}, A.~I., {et~al.} 2008, \apj,
  684, 1240

\bibitem[{{Fayolle} {et~al.}(2011){Fayolle}, {Bertin}, {Romanzin}, {Michaut},
  {{\"O}berg}, {Linnartz}, \& {Fillion}}]{fayolle2011}
{Fayolle}, E.~C., {Bertin}, M., {Romanzin}, C., {et~al.} 2011, \apjl, 739, L36

\bibitem[{{Fayolle} {et~al.}(2013){Fayolle}, {Bertin}, {Romanzin}, {Poderoso},
  {Philippe}, {Michaut}, {Jeseck}, {Linnartz}, {{\"O}berg}, \&
  {Fillion}}]{fayolle2013}
{Fayolle}, E.~C., {Bertin}, M., {Romanzin}, C., {et~al.} 2013, \aap, 556, A122

\bibitem[{{Frimann} {et~al.}(2016){Frimann}, {J{\o}rgensen}, \&
  {Haugb{\o}lle}}]{frimann2016}
{Frimann}, S., {J{\o}rgensen}, J.~K., \& {Haugb{\o}lle}, T. 2016, \aap, 587,
  A59

\bibitem[{{Furuya} {et~al.}(2015){Furuya}, {Aikawa}, {Hincelin}, {Hassel},
  {Bergin}, {Vasyunin}, \& {Herbst}}]{furuya2015}
{Furuya}, K., {Aikawa}, Y., {Hincelin}, U., {et~al.} 2015, \aap, 584, A124

\bibitem[{{Furuya} {et~al.}(2017){Furuya}, {Drozdovskaya}, {Visser}, {van
  Dishoeck}, {Walsh}, {Harsono}, {Hincelin}, \& {Taquet}}]{furuya2017}
{Furuya}, K., {Drozdovskaya}, M.~N., {Visser}, R., {et~al.} 2017, \aap, 599,
  A40

\bibitem[{{Furuya} {et~al.}(2016){Furuya}, {van Dishoeck}, \&
  {Aikawa}}]{furuya2016}
{Furuya}, K., {van Dishoeck}, E.~F., \& {Aikawa}, Y. 2016, \aap, 586, A127

\bibitem[{{Garrod}(2013)}]{garrod2013}
{Garrod}, R.~T. 2013, \apj, 765, 60

\bibitem[{{Garrod} \& {Herbst}(2006)}]{garrod2006}
{Garrod}, R.~T. \& {Herbst}, E. 2006, \aap, 457, 927

\bibitem[{{Garrod} \& {Pauly}(2011)}]{garrod2011}
{Garrod}, R.~T. \& {Pauly}, T. 2011, \apj, 735, 15

\bibitem[{{Garrod} {et~al.}(2007){Garrod}, {Wakelam}, \& {Herbst}}]{garrod2007}
{Garrod}, R.~T., {Wakelam}, V., \& {Herbst}, E. 2007, \aap, 467, 1103

\bibitem[{{Glover} {et~al.}(2010){Glover}, {Federrath}, {Mac Low}, \&
  {Klessen}}]{glover2010}
{Glover}, S.~C.~O., {Federrath}, C., {Mac Low}, M.~M., \& {Klessen}, R.~S.
  2010, \mnras, 404, 2

\bibitem[{{Goldsmith}(2001)}]{goldsmith2001}
{Goldsmith}, P.~F. 2001, \apj, 557, 736

\bibitem[{{Grassi} {et~al.}(2017){Grassi}, {Bovino}, {Haugb{\o}lle}, \&
  {Schleicher}}]{grassi2017}
{Grassi}, T., {Bovino}, S., {Haugb{\o}lle}, T., \& {Schleicher}, D.~R.~G. 2017,
  \mnras, 466, 1259

\bibitem[{{Grassi} {et~al.}(2014){Grassi}, {Bovino}, {Schleicher}, {Prieto},
  {Seifried}, {Simoncini}, \& {Gianturco}}]{grassi2014}
{Grassi}, T., {Bovino}, S., {Schleicher}, D.~R.~G., {et~al.} 2014, \mnras, 439,
  2386

\bibitem[{Harris {et~al.}(2020)Harris, Millman, van~der Walt, Gommers,
  Virtanen, Cournapeau, Wieser, Taylor, Berg, Smith, Kern, Picus, Hoyer, van
  Kerkwijk, Brett, Haldane, Fernández~del Río, Wiebe, Peterson,
  Gérard-Marchant, Sheppard, Reddy, Weckesser, Abbasi, Gohlke, \&
  Oliphant}]{2020NumPy-Array}
Harris, C.~R., Millman, K.~J., van~der Walt, S.~J., {et~al.} 2020, Nature, 585,
  357–362

\bibitem[{{Hasegawa} \& {Herbst}(1993)}]{hasegawa1993b}
{Hasegawa}, T.~I. \& {Herbst}, E. 1993, \mnras, 261, 83

\bibitem[{{Hasegawa} {et~al.}(1992){Hasegawa}, {Herbst}, \&
  {Leung}}]{hasegawa1992}
{Hasegawa}, T.~I., {Herbst}, E., \& {Leung}, C.~M. 1992, \apjs, 82, 167

\bibitem[{{Haugb{\o}lle} {et~al.}(2018){Haugb{\o}lle}, {Padoan}, \&
  {Nordlund}}]{haugbolle2018}
{Haugb{\o}lle}, T., {Padoan}, P., \& {Nordlund}, {\r{A}}. 2018, \apj, 854, 35

\bibitem[{{He} {et~al.}(2016){He}, {Acharyya}, \& {Vidali}}]{he2016}
{He}, J., {Acharyya}, K., \& {Vidali}, G. 2016, \apj, 823, 56

\bibitem[{{Heays} {et~al.}(2017){Heays}, {Bosman}, \& {van
  Dishoeck}}]{heays2017}
{Heays}, A.~N., {Bosman}, A.~D., \& {van Dishoeck}, E.~F. 2017, \aap, 602, A105

\bibitem[{{Heays} {et~al.}(2014){Heays}, {Visser}, {Gredel}, {Ubachs}, {Lewis},
  {Gibson}, \& {van Dishoeck}}]{heays2014}
{Heays}, A.~N., {Visser}, R., {Gredel}, R., {et~al.} 2014, \aap, 562, A61

\bibitem[{{Henning} \& {Semenov}(2013)}]{henning2013}
{Henning}, T. \& {Semenov}, D. 2013, Chemical Reviews, 113, 9016

\bibitem[{{Hocuk} {et~al.}(2017){Hocuk}, {Sz{\H{u}}cs}, {Caselli}, {Cazaux},
  {Spaans}, \& {Esplugues}}]{hocuk2017}
{Hocuk}, S., {Sz{\H{u}}cs}, L., {Caselli}, P., {et~al.} 2017, \aap, 604, A58

\bibitem[{{Hugo} {et~al.}(2009){Hugo}, {Asvany}, \& {Schlemmer}}]{hugo2009}
{Hugo}, E., {Asvany}, O., \& {Schlemmer}, S. 2009, \jcp, 130, 164302

\bibitem[{Hunter(2007)}]{hunter2007}
Hunter, J.~D. 2007, Computing in Science \& Engineering, 9, 90

\bibitem[{{Imai} {et~al.}(2019){Imai}, {Oya}, {Sakai}, {L{\'o}pez-Sepulcre},
  {Watanabe}, \& {Yamamoto}}]{imai2019}
{Imai}, M., {Oya}, Y., {Sakai}, N., {et~al.} 2019, \apj, 873, L21

\bibitem[{{Jensen} \& {Haugb{\o}lle}(2018)}]{jensen2018}
{Jensen}, S.~S. \& {Haugb{\o}lle}, T. 2018, \mnras, 474, 1176

\bibitem[{{Jensen} {et~al.}(2019){Jensen}, {J{\o}rgensen}, {Kristensen},
  {Furuya}, {Coutens}, {van Dishoeck}, {Harsono}, \& {Persson}}]{jensen2019}
{Jensen}, S.~S., {J{\o}rgensen}, J.~K., {Kristensen}, L.~E., {et~al.} 2019,
  \aap, 631, A25

\bibitem[{Jørgensen {et~al.}(2020)Jørgensen, Belloche, \&
  Garrod}]{jorgensen2020}
Jørgensen, J.~K., Belloche, A., \& Garrod, R.~T. 2020, Annual Review of
  Astronomy and Astrophysics, 58, null

\bibitem[{{Keto} \& {Caselli}(2008)}]{keto2008}
{Keto}, E. \& {Caselli}, P. 2008, \apj, 683, 238

\bibitem[{{Krumholz}(2014)}]{krumholz2014}
{Krumholz}, M.~R. 2014, \physrep, 539, 49

\bibitem[{{Kuffmeier} {et~al.}(2018){Kuffmeier}, {Frimann}, {Jensen}, \&
  {Haugb{\o}lle}}]{kuffmeier2018}
{Kuffmeier}, M., {Frimann}, S., {Jensen}, S.~S., \& {Haugb{\o}lle}, T. 2018,
  \mnras, 475, 2642

\bibitem[{{Kuffmeier} {et~al.}(2017){Kuffmeier}, {Haugb{\o}lle}, \&
  {Nordlund}}]{kuffmeier2017}
{Kuffmeier}, M., {Haugb{\o}lle}, T., \& {Nordlund}, {\r{A}}. 2017, \apj, 846, 7

\bibitem[{{Lee} {et~al.}(2019){Lee}, {Lee}, {Baek}, {Aikawa}, {Cieza}, {Yoon},
  {Herczeg}, {Johnstone}, \& {Casassus}}]{lee2019}
{Lee}, J.-E., {Lee}, S., {Baek}, G., {et~al.} 2019, Nature Astronomy, 3, 314

\bibitem[{{Leger} {et~al.}(1985){Leger}, {Jura}, \& {Omont}}]{leger1985}
{Leger}, A., {Jura}, M., \& {Omont}, A. 1985, \aap, 144, 147

\bibitem[{{Linsky}(2003)}]{linsky2003}
{Linsky}, J.~L. 2003, \ssr, 106, 49

\bibitem[{{Majumdar} {et~al.}(2017){Majumdar}, {Gratier}, {Ruaud}, {Wakelam},
  {Vastel}, {Sipil{\"a}}, {Hersant}, {Dutrey}, \& {Guilloteau}}]{majumdar2017}
{Majumdar}, L., {Gratier}, P., {Ruaud}, M., {et~al.} 2017, \mnras, 466, 4470

\bibitem[{{Masunaga} \& {Inutsuka}(2000)}]{masunaga2000}
{Masunaga}, H. \& {Inutsuka}, S.-i. 2000, \apj, 531, 350

\bibitem[{{Murillo} {et~al.}(2013){Murillo}, {Lai}, {Bruderer}, {Harsono}, \&
  {van Dishoeck}}]{murillo2013}
{Murillo}, N.~M., {Lai}, S.-P., {Bruderer}, S., {Harsono}, D., \& {van
  Dishoeck}, E.~F. 2013, \aap, 560, A103

\bibitem[{{Oberg} {et~al.}(2007){Oberg}, {Fuchs}, \& {van
  Dishoeck}}]{oberg2007}
{Oberg}, K.~I., {Fuchs}, G.~W., \& {van Dishoeck}, E.~F. 2007, in Molecules in
  Space and Laboratory, ed. J.~L. {Lemaire} \& F.~{Combes}, 80

\bibitem[{{{\"O}berg} {et~al.}(2009){{\"O}berg}, {van Dishoeck}, \&
  {Linnartz}}]{oberg2009}
{{\"O}berg}, K.~I., {van Dishoeck}, E.~F., \& {Linnartz}, H. 2009, \aap, 496,
  281

\bibitem[{{Pagani} {et~al.}(2007){Pagani}, {Bacmann}, {Cabrit}, \&
  {Vastel}}]{pagani2007}
{Pagani}, L., {Bacmann}, A., {Cabrit}, S., \& {Vastel}, C. 2007, \aap, 467, 179

\bibitem[{{Pagani} {et~al.}(1992){Pagani}, {Salez}, \& {Wannier}}]{pagani1992}
{Pagani}, L., {Salez}, M., \& {Wannier}, P.~G. 1992, \aap, 258, 479

\bibitem[{{Parise} {et~al.}(2005){Parise}, {Caux}, {Castets}, {Ceccarelli},
  {Loinard}, {Tielens}, {Bacmann}, {Cazaux}, {Comito}, {Helmich}, {Kahane},
  {Schilke}, {van Dishoeck}, {Wakelam}, \& {Walters}}]{parise2005}
{Parise}, B., {Caux}, E., {Castets}, A., {et~al.} 2005, \aap, 431, 547

\bibitem[{{Persson} {et~al.}(2014){Persson}, {J{\o}rgensen}, {van Dishoeck}, \&
  {Harsono}}]{persson2014}
{Persson}, M.~V., {J{\o}rgensen}, J.~K., {van Dishoeck}, E.~F., \& {Harsono},
  D. 2014, \aap, 563, A74

\bibitem[{{Richings} {et~al.}(2014){Richings}, {Schaye}, \&
  {Oppenheimer}}]{richings2014}
{Richings}, A.~J., {Schaye}, J., \& {Oppenheimer}, B.~D. 2014, \mnras, 442,
  2780

\bibitem[{{Rodgers} \& {Charnley}(2002)}]{rodgers2002}
{Rodgers}, S.~D. \& {Charnley}, S.~B. 2002, \planss, 50, 1125

\bibitem[{{Ruaud} {et~al.}(2016){Ruaud}, {Wakelam}, \& {Hersant}}]{ruaud2016}
{Ruaud}, M., {Wakelam}, V., \& {Hersant}, F. 2016, \mnras, 459, 3756

\bibitem[{{Semenov} {et~al.}(2003){Semenov}, {Henning}, {Helling}, {Ilgner}, \&
  {Sedlmayr}}]{semenov2003}
{Semenov}, D., {Henning}, T., {Helling}, C., {Ilgner}, M., \& {Sedlmayr}, E.
  2003, \aap, 410, 611

\bibitem[{{Shingledecker} {et~al.}(2019){Shingledecker}, {Vasyunin}, {Herbst},
  \& {Caselli}}]{shingledecker2019}
{Shingledecker}, C.~N., {Vasyunin}, A., {Herbst}, E., \& {Caselli}, P. 2019,
  \apj, 876, 140

\bibitem[{{Sipil{\"a}} {et~al.}(2015){Sipil{\"a}}, {Caselli}, \&
  {Harju}}]{sipila2015}
{Sipil{\"a}}, O., {Caselli}, P., \& {Harju}, J. 2015, \aap, 578, A55

\bibitem[{{Taquet} {et~al.}(2013){Taquet}, {Peters}, {Kahane}, {Ceccarelli},
  {L{\'o}pez-Sepulcre}, {Toubin}, {Duflot}, \& {Wiesenfeld}}]{taquet2013model}
{Taquet}, V., {Peters}, P.~S., {Kahane}, C., {et~al.} 2013, \aap, 550, A127

\bibitem[{{Teyssier}(2002)}]{teyssier2002}
{Teyssier}, R. 2002, \aap, 385, 337

\bibitem[{{Tielens} \& {Allamandola}(1987)}]{tielens1987}
{Tielens}, A.~G.~G.~M. \& {Allamandola}, L.~J. 1987, {Composition, Structure,
  and Chemistry of Interstellar Dust}, ed. D.~J. {Hollenbach} \& J.~{Thronson},
  Harley~A., Vol. 134, 397

\bibitem[{{Tobin} {et~al.}(2012){Tobin}, {Hartmann}, {Chiang}, {Wilner},
  {Looney}, {Loinard}, {Calvet}, \& {D'Alessio}}]{tobin2012}
{Tobin}, J.~J., {Hartmann}, L., {Chiang}, H.-F., {et~al.} 2012, \nat, 492, 83

\bibitem[{{van Dishoeck} {et~al.}(2014){van Dishoeck}, {Bergin}, {Lis}, \&
  {Lunine}}]{dishoeck2014}
{van Dishoeck}, E.~F., {Bergin}, E.~A., {Lis}, D.~C., \& {Lunine}, J.~I. 2014,
  in Protostars and Planets VI, ed. H.~{Beuther}, R.~S. {Klessen}, C.~P.
  {Dullemond}, \& T.~{Henning}, 835

\bibitem[{{Vastel} {et~al.}(2004){Vastel}, {Phillips}, \&
  {Yoshida}}]{vastel2004}
{Vastel}, C., {Phillips}, T.~G., \& {Yoshida}, H. 2004, \apjl, 606, L127

\bibitem[{{Visser} {et~al.}(2011){Visser}, {Doty}, \& {van
  Dishoeck}}]{visser2011}
{Visser}, R., {Doty}, S.~D., \& {van Dishoeck}, E.~F. 2011, \aap, 534, A132

\bibitem[{{Visser} {et~al.}(2009){Visser}, {van Dishoeck}, \&
  {Black}}]{visser2009b}
{Visser}, R., {van Dishoeck}, E.~F., \& {Black}, J.~H. 2009, \aap, 503, 323

\bibitem[{{Wakelam} {et~al.}(2015){Wakelam}, {Loison}, {Herbst}, {Pavone},
  {Bergeat}, {B{\'e}roff}, {Chabot}, {Faure}, {Galli}, {Geppert}, {Gerlich},
  {Gratier}, {Harada}, {Hickson}, {Honvault}, {Klippenstein}, {Le Picard},
  {Nyman}, {Ruaud}, {Schlemmer}, {Sims}, {Talbi}, {Tennyson}, \&
  {Wester}}]{wakelam2015}
{Wakelam}, V., {Loison}, J.~C., {Herbst}, E., {et~al.} 2015, \apjs, 217, 20

\bibitem[{{Wakelam} {et~al.}(2017){Wakelam}, {Loison}, {Mereau}, \&
  {Ruaud}}]{wakelam2017}
{Wakelam}, V., {Loison}, J.~C., {Mereau}, R., \& {Ruaud}, M. 2017, Molecular
  Astrophysics, 6, 22

\bibitem[{{Ward-Thompson} {et~al.}(2007){Ward-Thompson}, {Andr{\'e}},
  {Crutcher}, {Johnstone}, {Onishi}, \& {Wilson}}]{ward-thompson2007}
{Ward-Thompson}, D., {Andr{\'e}}, P., {Crutcher}, R., {et~al.} 2007, in
  Protostars and Planets V, ed. B.~{Reipurth}, D.~{Jewitt}, \& K.~{Keil}, 33

\bibitem[{{Wolcott-Green} \& {Haiman}(2011)}]{wolcott-green2011}
{Wolcott-Green}, J. \& {Haiman}, Z. 2011, \mnras, 412, 2603

\bibitem[{{Young} \& {Evans}(2005)}]{young2005}
{Young}, C.~H. \& {Evans}, Neal~J., I. 2005, \apj, 627, 293

\end{thebibliography}

\begin{appendix} 
\section{Surface reactions with custom barriers}\label{app:reactions}
\begin{table}
\caption{Reactions with custom barrier widths. Note that deuterated variants of the reactions have been excluded from the list, but share the same barrier widths.}
\centering
\begin{tabular}{lll}
\hline\hline
Reaction & Barrier width (\AA) & Reference \\\hline
O + CO $\rightarrow$ CO$_2$ & 1.25 & a,b  \\
H + CH$_4$ $\rightarrow$ CH$_3$ + H$_2$ & 2.17 & a,b \\
H + CO $\rightarrow$ HCO & 2.0 & b \\
H + H$_2$CO $\rightarrow$ CH$_2$OH & 2.0 & b \\
H + H$_2$CO $\rightarrow$ CH$_3$O & 2.0 & b \\
H + CH$_3$OH $\rightarrow$ CH$_2$OH + H$_2$ & 2.0 & b \\
H + CH$_3$OH $\rightarrow$ CH$_3$O + H$_2$ & 2.0 & b \\
\hline
\end{tabular}
\label{tab:reactions}
\tablebib{
(a) \citet{garrod2011}; (b) \citet{garrod2013}; (c) \citet{aikawa2012}}
\end{table}

\section{Comparison of D/H ratios toward 0.22 $\mathrm{M}_\odot$ protostars}\label{app:figures}

\begin{figure*}[ht]
\resizebox{\hsize}{!}
        {\includegraphics{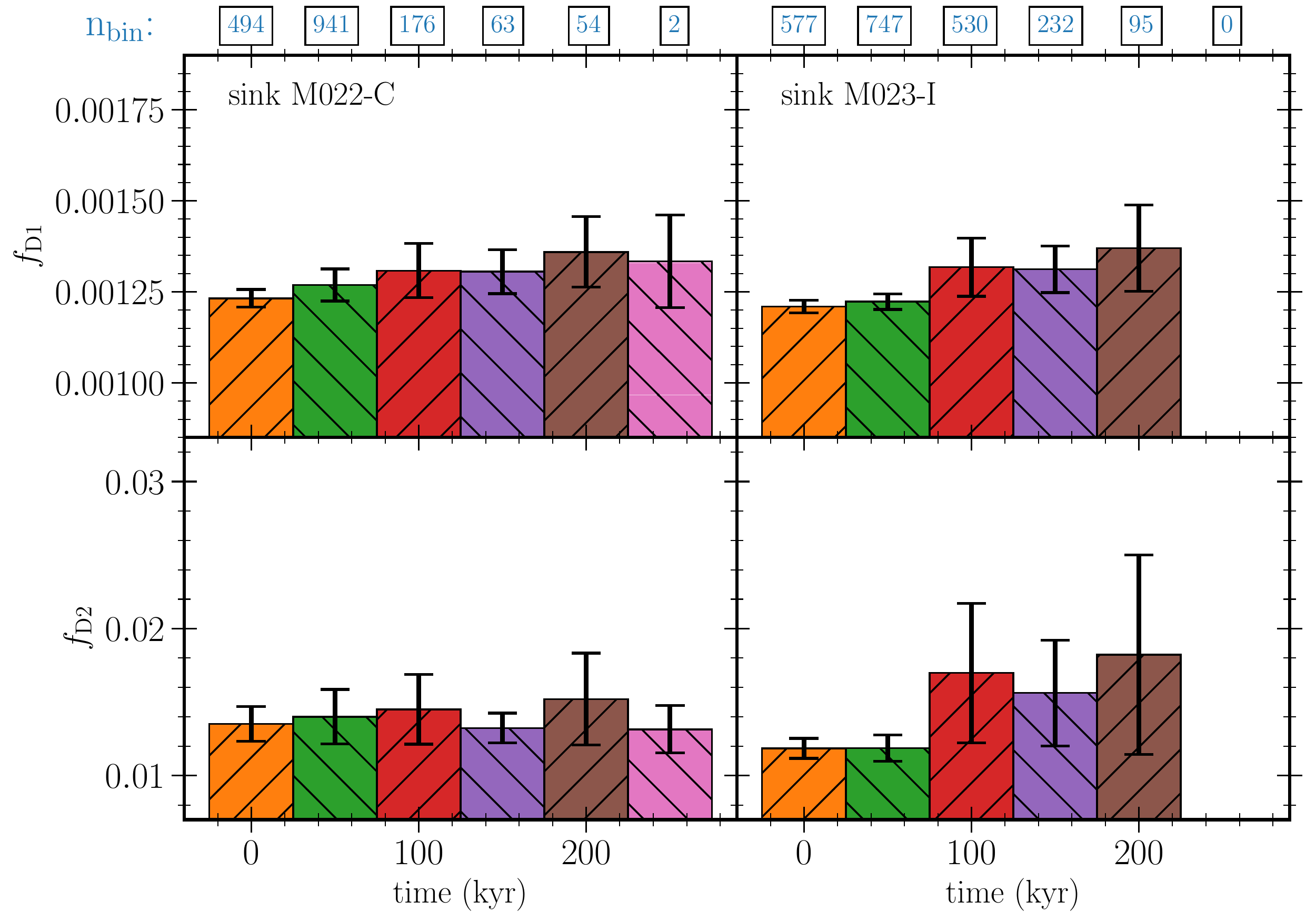}}
  \caption{HDO/H$_2$O ($f_{\mathrm{D}1}$) and D$_2$O/HDO ($f_{\mathrm{D}2}$) ratios toward three protostars in the simulation, with similar final mass of $\sim0.22 \mathrm{M}_\odot$. Tracer particles are binned according to the time at which they reach the hot corino, in bins with a width of 50 kyr. Time corresponds to the time after the onset of collapse, $t_0$. Each bar shows the median values of tracer particles accreted in the time interval, and the error shows the [15.9, 84.1] percentiles. The number of tracer particles within each bin is denoted in blue above the first row.}
     \label{fig:env_comparison3}
\end{figure*}

\section{Density structure of the cores at onset of collapse}\label{app:radial_profiles}
\begin{figure*}[ht]
\resizebox{\hsize}{!}
        {\includegraphics{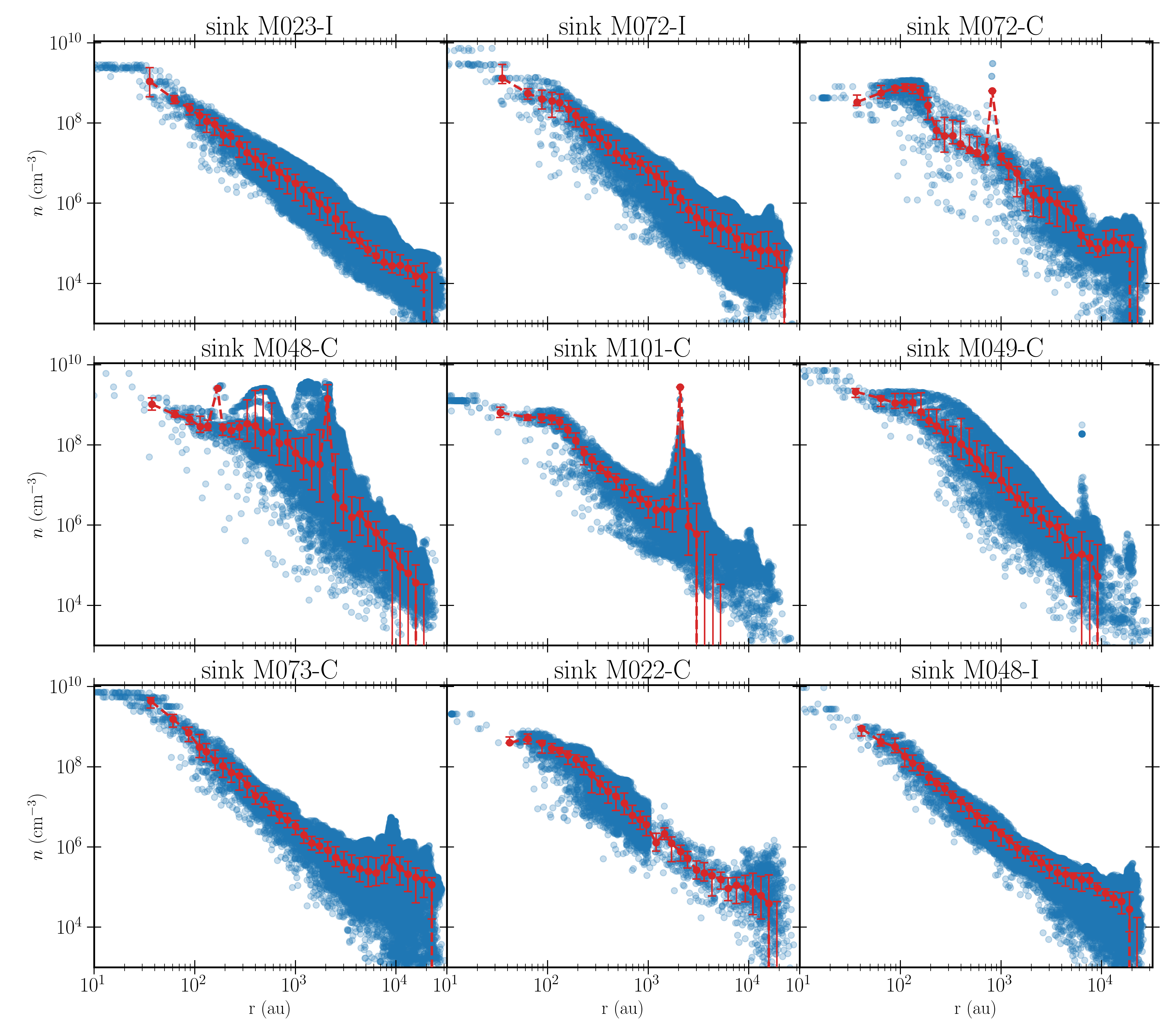}}
  \caption{Radial density profile for 40,000 tracer particle for the sinks studied in this work. The red line shows the median in each radial bin, while the errorbars show the 25 and 75 percentiles for each bin. Blue circles indicates individual tracer particles.}
     \label{fig:radial_density}
\end{figure*}

Figure \ref{fig:radial_density} shows the radial density profile for the 9 protostars studied in this work at the onset of collapse. The three isolated sink particles show a smooth Bonner-Ebert-like profile with $\rho \propto r^{-2}$. The clustered sources also resemble a Bonner-Ebert profile, however several sinks show indications of additional over-densities within 10,000~au, e.g., sink M048-C and sink M101-C. Overall, the profiles of the isolated cores also exhibit slightly less scatter around the medians.

\section{Lowering the ambient extinction $A_\mathrm{v}^{0}$}\label{app:av}
\begin{figure*}[ht]
\resizebox{\hsize}{!}
        {\includegraphics{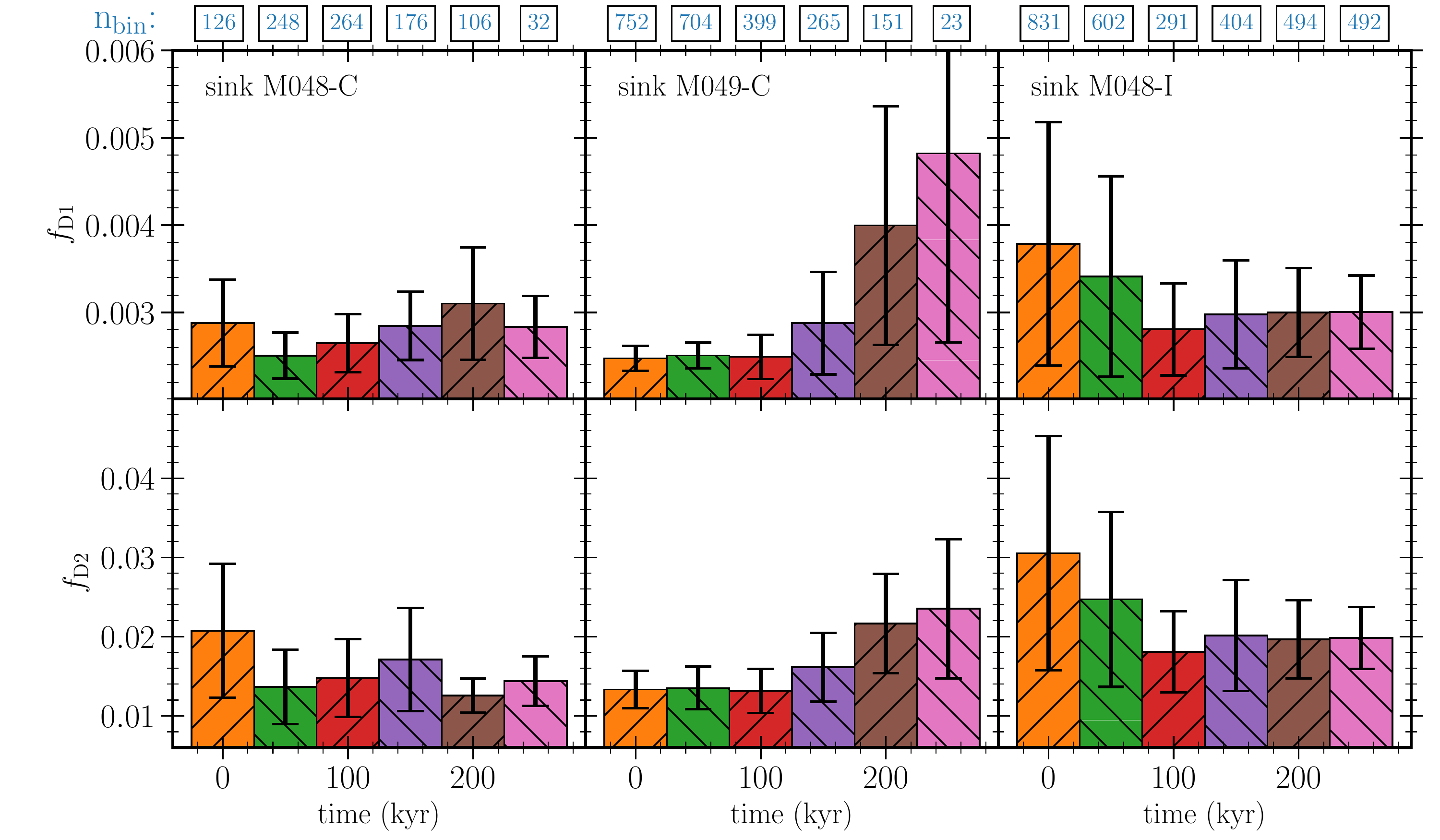}}
  \caption{Similar to Fig. \ref{fig:env_comparison}, with an ambient cloud extinction of 2~mag. HDO/H$_2$O ($f_{\mathrm{D}1}$) and D$_2$O/HDO ($f_{\mathrm{D}2}$) ratios toward three protostars in the simulation, with similar final mass of $\sim0.5 \mathrm{M}_\odot$. Tracer particles are binned according to the time at which they reach the hot corino, in bins with a width of 50 kyr. Time corresponds to the time after the onset of collapse, $t_0$. The third protostar, sink M048-I, is classified as \emph{isolated} since no protostar enter within 20,000~au during the simulation, while sink M048-C and sink M049-C are clustered. Each bar shows the median values of tracer particles accreted in the time interval, and the error shows the [15.9, 84.1] percentiles. The number of tracer particles within each bin is denoted in blue above the first row}
     \label{fig:env_comparison_av1}
\end{figure*}

\begin{figure*}[ht]
\resizebox{\hsize}{!}
        {\includegraphics{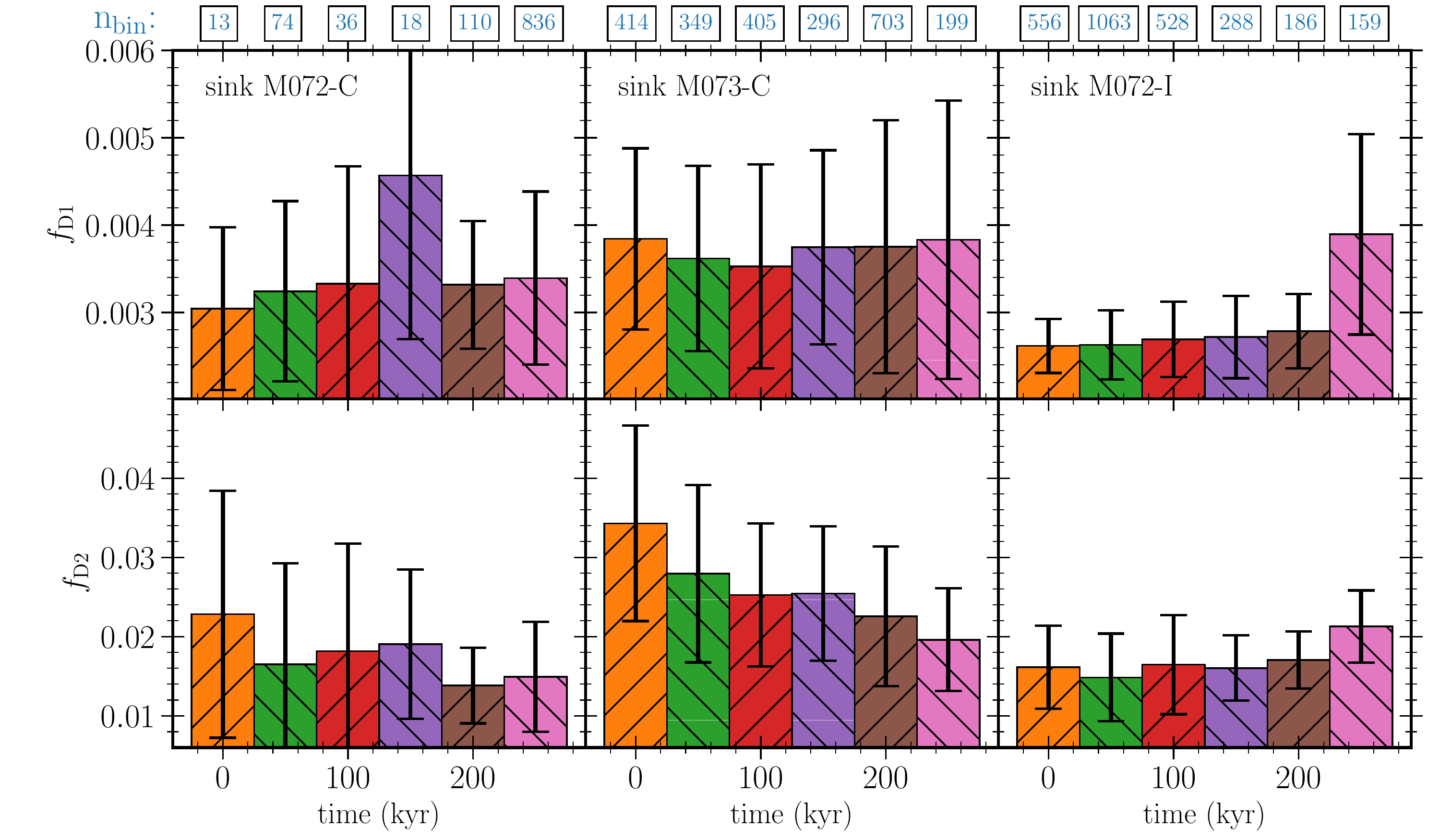}}
  \caption{Similar to Fig. \ref{fig:env_comparison2}, with an ambient cloud extinction of 2~mag. HDO/H$_2$O ($f_{\mathrm{D}1}$) and D$_2$O/HDO ($f_{\mathrm{D}2}$) ratios toward three protostars in the simulation, with similar final mass of $\sim0.7 \mathrm{M}_\odot$. Tracer particles are binned according to the time at which they reach the hot corino, in bins with a width of 50 kyr. Time corresponds to the time after the onset of collapse, $t_0$. Each bar shows the median values of tracer particles accreted in the time interval, and the error shows the [15.9, 84.1] percentiles. The number of tracer particles within each bin is denoted in blue above the first row}
     \label{fig:env_comparison_av2}
\end{figure*}

In the fiducial model, the ambient cloud is assumed to shield the cores with an extinction of 5~mag. To determine the impact of a lower ambient extinction, models with an ambient extinction of A$_\mathrm{v}^{0} = 2$~mag are presented here for 6 of the 9 protostars. Physical and chemical parameters are similar to the fiducial model, i.e., $T = 10$~K, $n_\mathrm{tot} = 2\times10^{4}$~cm$^{-3}$, and a duration of 3~Myr. In this case, the D/H ratios of water in the hot corino are slightly higher, and a larger spread in D/H ratios is observed. 
Figures \ref{fig:env_comparison_av1} and \ref{fig:env_comparison_av2} show the comparison between isolated and clustered protostars with similar final masses. No apparent correlation with the environment in present.


\end{appendix}
\end{document}